\documentclass[aps,superscriptaddress,showpacs,nofootinbib,eqsecnum]{revtex4}

\usepackage{epsfig}
\usepackage{latexsym}

\input{epsf}
\begin{document}
%auto-ignore

\title{Dynamics of Interacting Scalar Fields in Expanding 
Space-Time}

\author{Arjun Berera}
\email{ab@ph.ed.ac.uk}
\affiliation{School of Physics, University of 
Edinburgh, Edinburgh, EH9 3JZ, United Kingdom}

\author{Rudnei O. Ramos}
\email{rudnei@uerj.br}
\affiliation{Departamento de F\'{\i}sica Te\'orica,
Universidade do Estado do Rio de Janeiro,
20550-013 Rio de Janeiro, RJ, Brazil}

\begin{abstract}

The effective equation of motion is derived for a scalar field
interacting with other fields in a Friedman-Robertson-Walker 
background space-time.  The dissipative behavior reflected in this
effective evolution equation is studied both in simplified
approximations as well as numerically.
The relevance of our results to inflation are considered both in terms
of the evolution of the inflaton field as well as its fluctuation
spectrum.  A brief examination also is made of supersymmetric
models that yield dissipative effects during inflation.

\end{abstract}

\pacs{98.80.Cq}

\maketitle

\medskip

\centerline{\it Version published: Physical Review D71, 023513 (2005)}

\section{Introduction}
\label{intro}

It is by now well established that inflation models in general have
dissipative effects during the inflationary period
\cite{BGR,BGR2,BR,BR2,BR3}. These effects have two major consequences
for inflationary dynamics. First they result in radiation production
during inflation, which in turn influences the fluctuations that seed
large scale structure. Second they lead to temporally nonlocal terms in
the effective evolution equation of the inflaton, which can
significantly influence the nature and history of the inflation period.
The realization of these dissipative effects in inflation models has
resulted in the division of inflation into two dynamical possibilities
referred to as cold and warm inflation. Cold inflation is simply the
original picture of inflation envisioned in the earliest works
\cite{oldi,ni,ci}. In this picture the effects of dissipation are
negligible. The fluctuations created during inflation are effectively
zero-point ground state fluctuations and the evolution of the inflaton
field is governed by a ground state evolution equation. In contrast, in
the warm inflation picture \cite{wi}, inflationary expansion and radiation
production occur concurrently. In this picture, the fluctuations created
during inflation emerge from some excited statistical state and the
evolution of the inflaton has dissipative terms arising from the
interaction of the inflaton with other fields. 

The dividing point between warm and cold inflation is roughly at
$\rho_r^{1/4} \approx H$, where $\rho_r$ is the radiation energy density
present during inflation and $H$ is the Hubble parameter. Thus
$\rho_r^{1/4} > H$ is the warm inflation regime and $\rho_r^{1/4} < H$
is the cold inflation regime. This criteria is independent of
thermalization, but if such were to occur, one sees this criteria
basically amounts to the warm inflation regime corresponding to when $T
> H$. This is easy to understand since the typical inflaton mass during
inflation is $m_\phi \approx H$ and so when $T>H$, thermal fluctuations
of the inflaton field 
will become important. This criteria for entering the warm inflation
regime turns out to require the dissipation of a very tiny fraction of
the inflaton vacuum energy during inflation. {}For example, for
inflation with vacuum ({\it i.e.} potential) energy at the GUT scale
$\sim 10^{15-16} {\rm GeV}$, in order to produce radiation at the scale
of the Hubble parameter, which is $\approx 10^{10-11} {\rm GeV}$, it just
requires dissipating one part in $10^{20}$ of this vacuum energy density
into radiation. Thus energetically not a very significant amount of
radiation production is required to move into the warm inflation regime
\cite{HR}. In fact the levels are so small, and their eventual effects
on density perturbations and inflaton evolution are so significant, that
care must be taken to account for these effects in the analysis of any
inflation models.

In recent work \cite{BR2,BR3}, we have identified a key mechanism which
is generic in realistic inflation models and which leads to robust warm
inflation. This mechanism involves the scalar inflaton field $\phi$
exciting a heavy bosonic field $\chi$ which then decays to light
fermions $\psi_d$ \cite{BR},

\begin{equation}
\phi \rightarrow \chi \rightarrow \psi_d.
\label{wimech}
\end{equation}
In dynamical terms, this mechanism is expressed in its simplest form 
by an interaction Lagrangian density for the coupling of the
inflaton field to the other fields of the form

\begin{equation}
{\cal L}_I =  -\frac{1}{2}g^2 \phi^2 \chi^2 - 
g' \phi {\bar \psi_{\chi}} \psi_{\chi} - h \chi {\bar \psi_d}\psi_d ,
\label{lint}
\end{equation}

\noindent
where $\psi_d$ are the light fermions to which $\chi$-particles can
decay, with $m_\chi > 2m_{\psi_d}$. Aside from the last term in Eq.
(\ref{lint}), these are the typical interactions commonly used in
studies of reheating after inflation \cite{old,reheat}. However a
realistic inflation model often can also have additional interactions
outside the inflaton sector, with the inclusion of the light fermions
$\psi_d$ as depicted above being a viable option. Moreover in minimal
supersymmetry (SUSY) extensions of the typical reheating model or
multifield inflation models, the interactions of the form as given in
Eq. (\ref{lint}) can emerge as an automatic consequence of the
supersymmetric structure of the model. Since in the moderate to strong
perturbative regime, reheating and multifield inflation models will
require SUSY for controlling radiative corrections, Eq. (\ref{lint})
with inclusion of the $\psi_d$ field thus is a toy model representative
of many realistic situations.

In Ref. \cite{BR3} we have presented some results for the above
mechanism in a fully expanding space-time dissipative quantum field
theory formalism. The primary purpose of this paper is to supply the
full details of the formalism used in Ref. \cite{BR3}. This paper
presents the various approximations used in the derivation,
elaborates on the different aspects of dissipative evolution and
radiation production and discusses the impact of these results on the density
perturbation problem of inflation. Although the formalism presented here
has general application, for most of this paper we will focus on the
mechanism of Eq. (\ref{wimech}). The paper is organized as follows. In Sec.
\ref{compare} we compare the inflationary dissipative process
associated with Eq. (\ref{wimech})
to the reheating process after inflation in the standard cold inflation
models. A simple picture of dissipation for both processes is presented.
This should help to illustrate the differences between
the two processes at work in the two
cases. In Sec. \ref{model} the Lagrangian model studied in this paper is
presented. Also this section presents the real time matrix of Green's
functions needed for our calculations and their evolution equations in
expanding space-time. In Sec. \ref{solve} we present
approximate solutions for these evolution equations using the WKB {\it
ansatz}. The derivation of the effective equation of motion for the
inflaton field from a response theory approach is done in Sec.
\ref{eomderive}, where we also discuss several simplifying assumptions
for the nonlocal terms appearing in the effective equation of motion.
These basic equations are then studied numerically in Sec.
\ref{numerical} and the validity of the different approximation schemes
are explicitly tested for values of parameters of interest
to inflation. In Sec.
\ref{susy} we discuss SUSY models that 
realize the basic interaction structure
studied in this paper Eq. (\ref{lint}).  In Sec. \ref{iden} we study
the effects this dissipative process has on density perturbations 
during inflation.
Our concluding remarks are given in
Sec. \ref{concl}. An Appendix is also included where some details are
given on the renormalization of the effective equation of motion derived
in this paper.

\section{Interpretation of Dissipation in the linear and nonlinear regimes}

\label{compare}

It is useful to contrast the dissipation process to be discussed in the
following sections to the dissipation process in the
(old) reheating studies. {}For this, consider for instance the typical
models for reheating, where an inflaton field $\phi$, with potential
$V(\phi)$, is coupled either to spinor fields $\psi$ through the usual
Yukawa coupling $h \phi \psi \bar{\psi}$ and/or to other scalar fields
$\chi$, with coupling $g^2 \phi^2 \chi^2$, where in this last
case the inflaton potential has symmetry breaking, 
with minimum at $\phi = \phi_v$.

The typical reheating scenario is pictured in the time period just after
the inflationary regime, where the inflaton energy density is released
in the form of decay products of $\psi$ and/or $\chi$ particles. In the
reheating regime the Hubble constant $H$ is smaller than the inflaton
mass $m_\phi$, which means the inflaton can oscillate about the minimum,
$\phi_v$, of its potential $V(\phi)$. In addition, to have particle
production it requires $m_\phi$ to be sufficiently large so $\phi$ can
decay, $m_\phi > {\rm min}(2 m_\psi,2m_\chi)$. Typically one takes
$m_\phi \gg m_\psi,m_\chi$. In this regime the equation of motion is
simple to treat for small inflaton amplitude. For example in a quartic
inflaton potential with self-coupling $\lambda$, a perturbative
treatment is possible under the conditions $\lambda \phi(0) \ll m_\phi,
m_\chi, m_\psi$. In such a regime, the equation of motion for the
homogeneous inflaton field $\phi(t)$, including quantum corrections, is
given by the general linearized form \cite{old}

\begin{equation}
\ddot{\phi}(t) + 3 H \dot{\phi} (t) + \left[ m_\phi^2 + \Sigma(k) \right] 
\phi(t) =0\;,
\label{old-reh}
\end{equation}

\noindent
where $\Sigma(k)$ is the polarization, or self-energy operator for
$\phi$, with four-momentum $k=(\omega,0,0,0)$, with
$\omega=m_\phi$. Due to the condition $m_\phi > {\rm min}(2
m_\psi,2m_\chi)$ the $\phi$ self-energy has a nonzero imaginary part
${\rm Im \Sigma}$. The real part of the self-energy only renormalizes
the mass $m_\phi$, giving an effective mass to the inflaton, while the
imaginary part is associated with the damping of $\phi$ modes due to
decay, with the decay rate given by $\Gamma = - {\rm Im}
\Sigma(\omega)/(2\omega)$. In the regime ${\rm Im} \Sigma \ll m_\phi^2$,
the solution of (\ref{old-reh}) ends up being the same as if we just
replaced this equation with one having a friction like term proportional
to $\Gamma$, 

\begin{equation}
\ddot{\phi}(t) + 3 H \dot{\phi} (t) + m_\phi^2 \phi(t) + 
\Gamma \dot{\phi}(t) =0\;.
\label{old-reh2}
\end{equation}

\noindent
Note that in this derivation, since we are considering the regime
$H \ll m_\phi, m_\chi,m_\psi$,
the curvature of the universe is not important in the
calculation of the self-energy.  As such the decay
rate calculation is typically just done in Minkowski space-time.
Thus for the two decay processes of interest 
$\phi \to \chi+\chi$, or $\phi \to \psi + \bar{\psi}$, in
the rest frame of the $\phi$-particle the
respective decay rates are

\begin{eqnarray}
&&\Gamma_{\phi \to \chi \chi} = \frac{g^4 \phi_v^2}{8 \pi m_\phi}\;,
\nonumber\\
&& \Gamma_{\phi \to \psi \bar{\psi}} = \frac{h^2 m_\phi}{8 \pi}\;,
\label{srates}
\end{eqnarray}
where we have considered $m_\phi \gg m_\psi,m_\chi$.
Eq. (\ref{old-reh2}) with Eq, (\ref{srates}) comprise the basic
particle production process in the
old reheating studies.

\begin{figure}[ht]
\vspace{1cm}
\epsfysize=3cm
{\centerline{\epsfbox{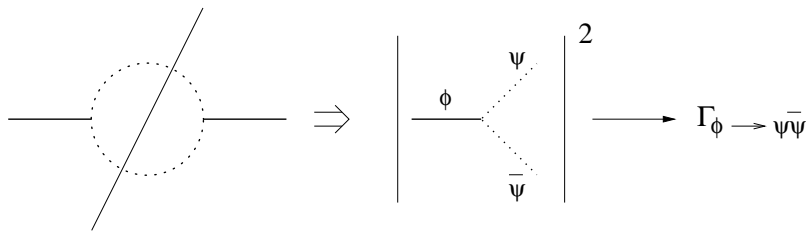}}}
\caption{The cut in the one-loop self-energy diagram and the amplitude of
the decay process associated to it. {}Full lines stand for the scalar $\phi$ 
and dotted lines to the spinors $\psi,\bar{\psi}$. Analogous process
follow for the decay $\phi \to \chi \chi$.}
\label{fig1}
\end{figure}

As is well known \cite{weldon}, the rates $\Gamma$ also can be expressed
in terms of the amplitude square for decay processes. This is
illustrated in {}Fig. \ref{fig1}, where the imaginary term contributing
to the rate of decays can be obtained by cutting the diagrams in half.
In a dynamical problem, when the $\phi$ modes of given energy $\omega$
and momentum ${\bf k}$ are displaced from equilibrium, these modes get
damped and it is the rate $\Gamma$ that describes the approach to
equilibrium. This approach to equilibrium is then naturally associated
to a dissipative, or irreversible process. This picture is therefore
closely related to that of a system (e.g. $\phi$) in interaction with an
environment (e.g. the $\chi$ bosons and $\psi$ fermions), in which
dissipation (and also noise) results from these interactions. More
formally stated in quantum mechanical terms \cite{weiss}, we describe
the state of the system by its reduced density matrix (by integrating out the
bath degrees). Due to interactions, the state of the system gets
entangled with the state of the bath and therefore some initial pure
state of the system will end up turning into a mixture. This is an
irreversible process, often called decoherence, that results from the
non-unitary system evolution.
 
We should also point out that although dissipation appears as a generic
consequence for a system in interaction with a bath or environment, the
simple representation for the dissipation, as in Eq. (\ref{old-reh2})
and the derivation of it, implies a number of simplifying assumptions
whose validity need to be checked. For instance, the derivation of Eq.
(\ref{old-reh2}) refers just to a very particular regime for the
inflaton field, when it is oscillating around the minimum of its
potential, with small field amplitudes and in the perturbative regime
(or the linear relaxation regime). For this case the effective equation
of motion of $\phi$ has the simplified linearized form Eq.
(\ref{old-reh2}). This simplified equation would not apply in nonlinear
regimes, when large field amplitudes dominate the dynamics, for example
in the description for the field modes during preheating, or in any
other situation where nonperturbative effects play a relevant role in
the description of field dynamics. Even in the linear regimes, Eq.
(\ref{old-reh2}) can be shown to be valid only up to a time interval
$\Delta t \lesssim 1/\Gamma$ \cite{boya}, beyond which the decay of
$\phi$ is no longer exponential but power law, which itself indicates
the break down of the perturbative approximation used to derive Eq.
(\ref{old-reh2}). It is clear that in more general cases of large field
amplitudes or beyond the perturbative approximation, the expected
effective equation of motion for the scalar field $\phi$ must become
very different than the simple equation (\ref{old-reh2}). Indeed, in
general the effective equation of motion for an arbitrary scalar
background field is a nonlocal equation. For example, in the nonlinear
regime or for high field amplitudes, the description of the effective
dynamics is not a simple local equation of motion (see e.g. Refs.
\cite{GR,BGR,BR}). In addition, as we move away from the regime of
validity of linear relaxation dynamics, it may become possible to find
other dissipative mechanisms not directly associated to the direct decay
process that leads to Eq. (\ref{old-reh2}). In fact, as shown in recent
work \cite{BR,BR2}, even in the case where the inflaton can not decay,
but fields coupled to it do (actually, $\phi$ does not even need to be
the heaviest field), dissipative regimes arise that are not available in
the linear or perturbative regime. 

\begin{figure}[ht]
\vspace{1cm}
\epsfysize=1.5cm
{\centerline{\epsfbox{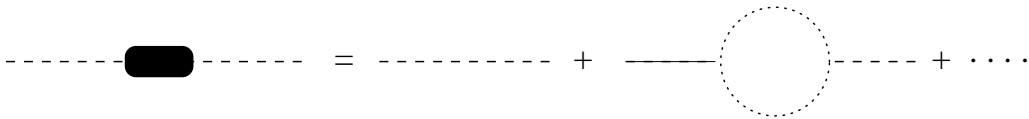}}}
\caption{The lowest order terms of the effective propagator for the scalar 
$\chi$ after integrating over the fermion fields coupled to it.
The black ellipse stand for the spinors quantum corrections 
and the dashed lines to the $\chi$ propagator.}
\label{fig2}
\end{figure}

\begin{figure}[ht]
\vspace{1cm}
\epsfysize=2.5cm
{\centerline{\epsfbox{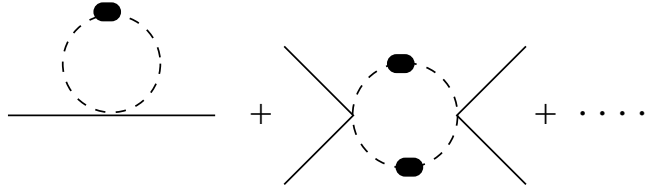}}}
\caption{Some of the lowest order diagrams contributing to the action of $\phi$
after integrating over the scalar $\chi$ and spinors $\psi_d,\bar{\psi}_d$.}
\label{fig3}
\end{figure}

\begin{figure}[ht]
\vspace{1cm}
\epsfysize=3cm
{\centerline{\epsfbox{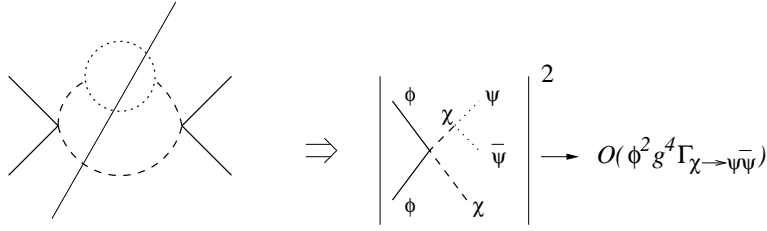}}}
\caption{The cut in the lowest order nontrivial diagram
appearing in the effective action of $\phi$ that has an imaginary part and
the main amplitude associated to it.}
\label{fig4}
\end{figure}

An example of interactions leading to a nontrivial 
inflaton dissipative dynamics
is for instance the ones shown in Eq.
(\ref{lint}), where now $m_\chi > {\rm min}(2 m_{\psi_d},m_\phi)$ and
$m_\phi < {\rm min}( m_{\psi_\chi}, m_\chi)$. Here the $\chi$
particles can decay into fermions $\psi_d$ that are coupled to it, but
there are no kinematically allowed direct decays of $\phi$ into other
particles. Nevertheless it is  simple to understand
from elementary particle physics
the origin and nature of
dissipation for the inflaton in this case.
We look
at those processes involving $\phi$ that may have an imaginary term and
so in analogy to (\ref{old-reh}), can be associated to
dissipation. This is better interpreted in terms of an effective theory
for $\phi$ after integrating over the other fields. We can start doing
this by first integrating over the $\psi_d$ fermion field. 
Since this field only
couples to $\chi$, its main effect is to dress the $\chi$
scalar propagator,
as shown schematically in {}Fig. \ref{fig2}. Note that the lowest
order correction to $\chi$ goes exactly like the previous case analyzed
above, in which the $\phi$ field 
could decay into $\psi$ particles. The leading
order one-loop self-energy contribution to $\chi$, $\Sigma_\chi (k)$,
has a real part that represents a shift in the mass squared of the
$\chi$ and an imaginary part that represents the rate of its decay, as
is kinematically allowed. This is the same kind of process shown in
{}Fig. \ref{fig1}, by replacing the external lines by $\chi$ and the
internal ones by the spinors $\psi_d$. Next we can now perform the
integration over the (dressed) $\chi$ scalar particles and the
$\psi_{\chi}$ spinors coupled to $\phi$. The relevant contributions to
our dissipative mechanism are due to 
the $\chi$ decay (see, however, later in
Sec. V a discussion on the role of the $\psi_{\chi}$ spinors). We now
have processes contributing to the effective action to $\phi$ like the
ones shown in {}Fig. \ref{fig3}. At leading order, the important
contribution to dissipation in the effective equation of motion for
$\phi$ arises now from the one-loop vertex diagram shown in {}Fig.
\ref{fig4}. By cutting that diagram in half we are now led to an
imaginary contribution that can be seen as a dissipative term appearing
in the effective equation of motion of $\phi$. This can be understood
from the amplitude shown in {}Fig. \ref{fig4}, which represents a
scattering process of a $\phi$ by a virtual $\chi$ that then decays into
the fermion particles. This process can be interpreted in terms of the
effective theory for $\phi$, where an evolving background $\phi$-field
configuration excites $\phi$-energy modes which then decay into the
light fermions $\psi_d$, with that decay mediated by (virtual) $\chi$
particles. The resulting dissipative term appearing in the effective
equation of motion for $\phi$ can be seen from the square amplitude
shown in {}Fig. \ref{fig4} to be of order ${\cal O}(\phi^2 g^4
\Gamma_{\chi\to \psi_d \bar{\psi}_d})$. This result is in fact
corroborated by the explicit derivation of the dissipation term in
Sec. V. We can also note that this result is nonlinear in the $\phi$
field amplitude, since it originates from a scattering process
(involving two $\phi$ particles and two virtual $\chi$ particles) and it
is nonperturbative in nature, since it involves the dressed $\chi$
propagators based on a resummation, which means that physically the
$\phi$ field is not interacting with vacuum like $\chi$ excitations but rather
with the collective $\chi$ excitations.

It should be noted that the dissipative processes as represented in
{}Fig. \ref{fig4} are of higher order than the ones shown in {}Fig.
\ref{fig1}. However, due to the nontrivial nature of the latter, they
may become important in those regimes characterized by high $\phi$
amplitudes (and then outside the region of validity of linear relaxation
theory), and couplings ($g$ and $h$ in Eq. (\ref{lint})) that are not
perturbatively small. It is exactly in this region of parameters, that we
will find the relevance to inflationary dynamics of the nonlinear
dissipative mechanism discussed here and derived in this paper. 

\section{A working model of a scalar field in interaction with 
multiple other fields}
\label{model}

Consider the following model initially presented in \cite{BR2}, which
consists of a scalar field $\Phi$ interacting with a set of scalar
fields $\chi_j$, $j=1, \ldots ,N_\chi$ and these scalar fields in turn
interact with fermion fields $\psi_k$, $k=1, \ldots, N_\psi$. Here we
work in the FRW background metric $ds^2=dt^2 -a(t)^2 d{\bf x}^2$. The
Lagrangian density for the matter fields coupled to the gravitational
field tensor $g_{\mu \nu}$ is given by 
 
\begin{eqnarray} 
{\cal L} [ \Phi, \chi_j, \bar{\psi}_k, \psi_k, g_{\mu \nu}] &=&  
\sqrt{-g} \left\{ \frac{1}{2} g^{\mu \nu} 
\partial_\mu \Phi \partial_\nu \Phi - \frac{m_\phi^2}{2}\Phi^2 - 
\frac{\lambda}{4 !} \Phi^4  -\frac{\xi}{2} R \Phi^2
\right. \nonumber \\
&+& \left. \sum_{j=1}^{N_{\chi}} \left[ 
g^{\mu \nu}  \frac{1}{2} \partial_\mu \chi_{j} \partial_\nu \chi_{j} - 
\frac{m_{\chi_j}^2}{2}\chi_j^2 
- \frac{f_{j}}{4!} \chi_{j}^4 - \frac{g_{j}^2}{2} 
\Phi^2 \chi_{j}^2  -\frac{\xi}{2} R \chi_j^2
\right] \right.
\nonumber \\ 
&+& \left. \sum_{k=1}^{N_{\psi}}   \left[
i \psi_{k} \gamma^\mu \left(\partial_\mu + \omega_\mu \right)\psi_k - 
\bar{\psi}_k\left( m_{\psi_k}  
+ \sum_{j=1}^{N_\chi} h_{kj,\chi} \chi_j \right)  \psi_k \right] \right\} 
\: , 
\label{Nfields} 
\end{eqnarray} 
 
\noindent  
where $R$ is the curvature scalar, 

\begin{equation}
R= 6 \left(\frac{\ddot{a}}{a} + \frac{\dot{a}^2}{a^2} \right) \;,
\end{equation}
$\xi$ is the dimensionless parameter describing the coupling of the
matter fields to the gravitational background and all coupling constants
are positive: $\lambda$, $f_{j},g_{j}^2, h_{kj,\chi}$ $> 0$. In the last
term involving the fermion fields, the $\gamma^\mu$ matrices are related
to the vierbein $e_\mu^a$ (where $g_{\mu\nu}=e_\mu^a e_\nu^b \eta_{ab}$,
with $\eta_{ab}$ the usual Minkowski metric tensor) by \cite{mallik}
$\gamma^\mu (x) = \gamma^a e_a^\mu (x)$, where $\gamma^a$ are the usual
Dirac matrices and $\omega_\mu = -(i/4) \sigma^{ab} e_a^\nu \nabla_\mu
e_{b\nu}$, with $\sigma^{ab} =i [\gamma^a,\gamma^b]/2$.
  
We are interested in obtaining the effective equation of motion (EOM)
for a scalar field configuration $\varphi = \langle \Phi \rangle$ after
integrating out the $\Phi$ fluctuations, the scalars $\chi_j$ and
spinors $\psi_k,\bar{\psi}_k$. This is a typical ''system-environment''
decomposition of the problem in which $\varphi$ is regarded as the
system field and everything else is the environment, which in particular
means the $\Phi$ fluctuation modes, the scalars $\chi_j$ and the spinors
$\psi_k,\bar{\psi}_k$ are regarded as the environment bath. In a
Minkowski background, at $T=0$, the EOM for $\varphi$ has been derived
in \cite{BR} using the Schwinger closed time path formalism. Here we
follow a completely analogous approach and derive the EOM in a FRW
background. The field equation for $\Phi$ can be readily obtained from
Eq. (\ref{Nfields}) and it is given by

\begin{eqnarray} 
&&\ddot{\Phi} + 3 \frac{\dot{a}}{a} \dot{\Phi}
-\frac{\nabla^2}{a^2} \Phi + m_\phi^2 \Phi + \frac{\lambda}{6} \Phi^3
+\xi R \Phi  
+\sum_{j=1}^{N_{\chi}} g_j^2 \Phi (x) \chi_j^2(x)  = 0 \;. 
\label{eqphi1} 
\end{eqnarray} 
In order to obtain the effective EOM for $\varphi$, we use the
tadpole method.  In this
method we split $\Phi$ in Eq.
(\ref{eqphi1}), as usual, into the (homogeneous) classical expectation
value $\varphi(t) =\langle \Phi \rangle$ and a quantum fluctuation
$\phi(x)$, $\Phi(x) = \varphi(t) + \phi(x)$. This way, the field equation for
$\Phi$, after taking the average (with $\langle \phi (x) \rangle=0$),
becomes
    
\begin{eqnarray} 
&&\ddot{\varphi}(t) + 3 \frac{\dot{a}(t)}{a(t)} \dot{\varphi}(t)+ 
m_\phi^2 \varphi(t) + 
\frac{\lambda}{6} \varphi^3(t)  +\xi R(t) \varphi(t)  
+\frac{\lambda}{2} \varphi(t) \langle \phi^2 \rangle 
+\frac{\lambda}{6} \langle \phi^3 \rangle \nonumber \\
&+& \sum_{j=1}^{N_{\chi}} g_j^2 \left[\varphi (t) \langle \chi_j^2 \rangle + 
\langle \phi \chi_j^2 \rangle \right] = 0 \;, 
\label{eqphi2} 
\end{eqnarray}

\noindent 
where $\langle \phi^2 \rangle$, $\langle \phi^3 \rangle$, $\langle
\chi_j^2 \rangle$ and $\langle \phi \chi_j^2 \rangle$ can be
expressed \cite{BR} in terms
of the coincidence limit of the (causal) two-point Green's functions
$G^{++}_\phi (x,x')$ and $G^{++}_{\chi_j} (x,x')$, for the $\Phi$ and
$\chi_j$ fields respectively. These Green's functions are appropriately 
defined in the context of the Schwinger closed time path (CTP) formalism
\cite{CTP}.  They are obtained from
the $(1,1)$-component of the real time matrix of effective propagators, 
which
satisfy the appropriate Schwinger-Dyson equations (see, {\it e.g.},
\cite{ian,BGR} for further details).
These equations satisfied by the effective (or dressed) propagators
emerge from the successive integrations over the bath fields in
(\ref{Nfields}). The integration over the spinors $\psi_k,\bar{\psi}_k$ 
leads to dressed propagators for the $\chi_j$ fields (see
e.g. {}Fig. \ref{fig2}), which are then given by
(in the {}FRW background)

\begin{eqnarray}
\left[\frac{\partial^2}{\partial t^2} + 3 \frac{\dot{a}}{a} 
\frac{\partial}{\partial t}
-\frac{\nabla^2}{a^2} + m_{\chi_j}^2 + g_j^2
\varphi(t)^2 + \xi R(t) \right] 
G_{\chi_j} (x,x')
+ \int d^4 z \Sigma_{\chi_j} (x,z) G_{\chi_j} (z,x') = i
\frac{\delta (x,x')}{a^{3/2}(t) a^{3/2}(t')} \;,
\label{Gchi}
\end{eqnarray}

\noindent
where $\Sigma_{\chi_j} (x,y)$ is the self-energy for $\chi_j$
due to the coupling to the spinors $\psi_k,\bar{\psi}_k$.
Next, by integrating over the $\chi_j$ and $\phi$ fluctuations
we are left with an effective propagator for the $\Phi$ fields
that is also formally defined by

\begin{eqnarray}
\left[\frac{\partial^2}{\partial t^2} + 3 \frac{\dot{a}}{a} 
\frac{\partial}{\partial t}
-\frac{\nabla^2}{a^2} + m_\phi^2 + \frac{\lambda}{2}
\varphi(t)^2 + \xi R(t) \right] 
G_\phi (x,x')
+ \int d^4 z \Sigma_\phi (x,z) G_\phi (z,x') = i 
\frac{\delta (x,x')}{a^{3/2}(t) a^{3/2}(t')}  \;,
\label{Gphi}
\end{eqnarray}

\noindent
where $\Sigma_\phi (x,y)$ denotes now the self-energy for
$\Phi$ after integrating over the remaining bath fields.

In the CTP formalism of quantum field theory, Eqs. (\ref{Gchi})
and (\ref{Gphi}) are matrix equations for the propagators
and self-energies, where, for instance, by expressing
$G_\phi (x,x')$ in terms of its
momentum-space {}Fourier transform (and
analogously for $G_{\chi_j}$), it can be expressed in the form

\begin{equation}
G_\phi(x,x') =i  \int \frac{d^3 q}{(2 \pi)^3} 
e^{i {\bf q} . ({\bf x} - {\bf x}')}
\left(
\begin{array}{ll}
G^{++}_\phi({\bf q}, t,t') & \:\: G^{+-}_\phi({\bf q}, t,t') \\
G^{-+}_\phi({\bf q}, t,t') & \:\: G^{--}_\phi({\bf q}, t,t')
\end{array}
\right) \: ,
\label{Gmatrix}
\end{equation}

\noindent
where

\begin{eqnarray}
&& G^{++}_\phi({\bf q} , t,t') = G^{>}_\phi({\bf q},t,t')
\theta(t-t') + G^{<}_\phi({\bf q},t,t') \theta(t'-t) ,
\nonumber \\
& & G^{--}_\phi({\bf q} , t,t') = G^{>}_\phi({\bf q},t,t')
\theta(t'-t) + G^{<}_\phi({\bf q},t,t') \theta(t-t') ,
\nonumber \\
& & G^{+-}_\phi({\bf q} , t,t') = G^{<}_\phi({\bf q},t,t') ,
\nonumber \\
& & G^{-+}_\phi({\bf q},t,t') = G^{>}_\phi({\bf q},t,t')\; ,
\label{G of k}
\end{eqnarray}

\noindent
with each matrix element defined in terms of two-point correlations
of the fields that are on each branch of the CTP contour \cite{CTP}.
In writing (\ref{Gmatrix}) we are assuming that $G(x,x')$
only depends on the difference ${\bf x}-{\bf x}'$, which follows
for homogeneous field configurations, as is our interest here.
The elements of the propagator matrix, Eq. (\ref{G of k}), for
both the scalars bosons $\phi$ and $\chi_j$, are 
found to satisfy the conditions

\begin{eqnarray}
&& G^>(x,x') =G^<(x',x)\;,\nonumber \\
&& \left[iG^{>(<)}(x,x')\right]^\dagger  = iG^{<(>)} (x,x') \;,\nonumber \\
&&\left[G^{>(<)} ({\bf q},t,t')\right]^\dagger = G^{>(<)}({\bf q},t',t)\;,
\nonumber \\
&&\frac{d}{dt}
\left[G^{>}({\bf q},t,t') - 
G^{>}({\bf q},t',t)\right]\Bigr|_{t=t'}=i \delta(t-t')\;,
\label{cond}
\end{eqnarray}

\noindent
where the third condition in (\ref{cond}) is just a result
of the first two conditions. The definitions of the retarded and
advanced propagators are also given in terms of the matrix
elements of the two-point function in the CTP formalism:

\begin{eqnarray}
&&G^{\rm ret}(x,x')= \theta(t-t') \left[
G^>(x,x') - G^{<}(x,x') \right]=
G^{++}(x,x') - G^{+-}(x,x')\;,
\label{ret}
\\
&&G^{\rm adv}(x,x')= \theta(t'-t) \left[
G^<(x,x') - G^{>}(x,x') \right]=
G^{++}(x,x') - G^{-+}(x,x')\; .
\label{adv}
\end{eqnarray}
In particular, we have the known result that
$G^{\rm ret}(x,x')=G^{\rm adv}(x',x)$.

The self-energy matrix elements are also expressed in a similar way
to the propagators in the CTP formalism as

\begin{eqnarray}
&& \Sigma_\phi^{++}(x,x') = \Sigma_\phi^{>}(x,x')
\theta(t-t') + \Sigma_\phi^{<}(x,x') \theta(t'-t) ,
\nonumber \\
& & \Sigma_\phi^{--}(x,x') = \Sigma_\phi^{>}(x,x')
\theta(t'-t) + \Sigma_\phi^{<}(x,x') \theta(t-t') ,
\nonumber \\
& & \Sigma_\phi^{+-}(x,x') =- \Sigma^{<}_\phi(x,x') ,
\nonumber \\
& & \Sigma^{-+}_\phi(x,x') = -\Sigma^{>}_\phi(x,x')\; ,
\label{Sigma}
\end{eqnarray}
with analogous expressions for the $\chi_j$ self-energy elements.
{}From (\ref{Sigma}) the following property follows,

\begin{equation}
\Sigma^{++}(x,x') + \Sigma^{+-}(x,x') + \Sigma^{-+}(x,x') +
\Sigma^{--}(x,x') =0\;.
\label{sum Sigma}
\end{equation}

\noindent
In addition the elements in (\ref{Sigma}) are
found in general to satisfy conditions similar to those
in (\ref{cond}), which are
valid for both the scalar bosons $\phi$ and $\chi_j$,

\begin{eqnarray}
&&\Sigma^{>}(x,x') = \Sigma^< (x',x)\;,
\nonumber \\
&&
\left[i\Sigma^{>(<)}(x,x')\right]^\dagger  = i\Sigma^{<(>)} (x,x') \;.
\label{Sigma cond}
\end{eqnarray}

In terms of Eq. (\ref{Sigma}) and the equations for the fluctuation field modes
derived from Eq. (\ref{Nfields}), we can write general 
expressions for the solutions
for the $G^{>(<)}$ propagator functions. {}For instance, consider the fluctuations
equation for the $\chi_j$ fields that can be obtained after integrating over
the fermion $\psi_k,\bar{\psi}_k$. This equation is obtained from the quadratic
action in the $\chi_j$ scalar fields, $S_2 [\chi_j]$ obtained
from Eq. (\ref{Nfields}). In the CTP formalism it is obtained by 
identifying fields in each branch of the Schwinger's closed time path
contour, with fields in the forward and backward segments
of the CTP time contour identified as $\chi_j^{+}$ and
$\chi_j^{-}$, respectively (see e.g. \cite{GR}).
In term of these fields, we can express the quadratic action for
the $\chi_j$ fields, after integration over the fermions as

\begin{eqnarray}
S_2[\chi_j^{+},\chi_j^{-}] &=&  \frac{1}{2}  \int d^4 x a^3(t)
\left\{
\left[ \left(\frac{\partial \chi_j^{+}}{\partial t} \right)^2 -
\chi_j^{+} \left(-\frac{\nabla^2}{a^2(t)} + M_{\chi_j}^2 \right) \chi_j^{+}
\right]  -
\left[ \left(\frac{\partial \chi_j^{-}}{\partial t} \right)^2 -
\chi_j^{-} \left(-\frac{\nabla^2}{a^2(t)} + M_{\chi_j}^2 \right) \chi_j^{-}
\right] \right\}
\nonumber \\
&-& \int d^4x a^3(t) \int d^4x' a^3(t') \; \frac{1}{2}
\left[ \chi_j^{+} (x) \Sigma_{\chi_j}^{++} (x,x') \chi_j^{+}(x')
+ \chi_j^{+} (x) \Sigma_{\chi_j}^{+-} (x,x') \chi_j^{-}(x')
\right. \nonumber \\
&& \left. ~~~~~~~~~~~~~~~~~~~~~~~~~~~~~~~~+\chi_j^{-} (x) 
\Sigma_{\chi_j}^{-+} (x,x') \chi_j^{+}(x')
+ \chi_j^{-} (x) \Sigma_{\chi_j}^{--} (x,x') \chi_j^{-}(x')\right]\;,
\label{S+-}
\end{eqnarray}

\noindent
where $M_{\chi_j}^2 = m_{\chi_j}^2 + g_j^2 \varphi^2(t) +\xi R(t) $,
$\varphi(t)$ is the background $\Phi$ field,
and $\Sigma_{\chi_j}(x,x')$ denotes the fermion
loop contributions, which dress the $\chi_j$ fields.
It is now useful to use in Eq. (\ref{S+-}) the redefined fields
given by \cite{GR}

\begin{eqnarray}
&& \chi_j^{c} = \frac{1}{2} \left(\chi_j^{+} + \chi_j^{-} \right) \;,
\nonumber \\
&& \chi_j^{\Delta} = \chi_j^{+} - \chi_j^{-}  \;, 
\label{new chi}
\end{eqnarray}

\noindent
along with the identity (\ref{sum Sigma}), which lead to the
result

\begin{eqnarray}
S_2[\chi_j^{c},\chi_j^{\Delta}] &=&  \int d^4 x a^3(t)
\left[ \chi_j^{\Delta} \left(
-\frac{\partial^2}{\partial t^2} - 3 \frac{\dot{a}}{a} \frac{\partial}{\partial t}
+\frac{\nabla^2}{a^2} - M_{\chi_j}^2 \right) \chi_j^{c}
\right]
\nonumber \\
&-& \int d^4x a^3(t) \int d^4x' a^3(t')
\left\{ \chi_j^{\Delta} (x) \left[
\Sigma_{\chi_j}^{++} (x,x') + \Sigma_{\chi_j}^{+-} (x,x') -
\Sigma_{\chi_j}^{-+} (x,x') - \Sigma_{\chi_j}^{--} (x,x') \right]
\chi_j^{c} (x') \right.  \nonumber \\
&& \left. ~~~~~~~~~~~~~~~~~~~~~~~~~~~-\chi_j^{\Delta} (x) \left[
\Sigma_{\chi_j}^{++} (x,x') - \Sigma_{\chi_j}^{+-} (x,x') -
\Sigma_{\chi_j}^{-+} (x,x') + \Sigma_{\chi_j}^{--} (x,x') \right]
\chi_j^{\Delta} (x') \right\}\;.
\label{S2}
\end{eqnarray}

\noindent
Using Eqs. (\ref{Sigma}) and (\ref{sum Sigma}), the argument involving 
the self-energies in the second term in Eq. (\ref{S2}) becomes

\begin{eqnarray}
&& \Sigma_{\chi_j}^{++} (x,x') + \Sigma_{\chi_j}^{+-} (x,x') -
\Sigma_{\chi_j}^{-+} (x,x') - \Sigma_{\chi_j}^{--} (x,x') 
\nonumber \\
&&= 2 \left[ \Sigma_{\chi_j}^{++} (x,x') + \Sigma_{\chi_j}^{+-} (x,x') 
\right] = 2 \theta(t_1-t_2) \left[ \Sigma_{\chi_j}^{>} (x,x') - 
\Sigma_{\chi_j}^{<} (x,x') 
\right] \nonumber \\
&&= \Pi_{\chi_j} (x,x') =\Pi_{1,\chi_j}(x,x') + \Pi_{2,\chi_j} (x,x')\;,
\label{2nd}
\end{eqnarray}

\noindent
where

\begin{eqnarray}
&& \Pi_{1,\chi_j}(x,x') = \left[2 \theta(t_1-t_2) -1\right]
\left[  \Sigma_{\chi_j}^{>} (x,x') - 
\Sigma_{\chi_j}^{<} (x,x') 
\right] \;,
\nonumber \\
&& \Pi_{2,\chi_j}(x,x') = \Sigma_{\chi_j}^{>} (x,x') - 
\Sigma_{\chi_j}^{<} (x,x') \;,
\label{Pis}
\end{eqnarray}

\noindent
which, from (\ref{Sigma cond}) 
have the properties $\Pi_{1,\chi_j} (x,x') = \Pi_{1,\chi_j} (x',x)$
and $\Pi_{2,\chi_j} (x,x') = -\Pi_{2,\chi_j} (x',x)$.
Similarly, the self-energy contributions in the third term in
Eq. (\ref{S2}) can be written as

\begin{eqnarray}
&& \Sigma_{\chi_j}^{++} (x,x') - \Sigma_{\chi_j}^{+-} (x,x') -
\Sigma_{\chi_j}^{-+} (x,x') + \Sigma_{\chi_j}^{--} (x,x') 
\nonumber \\
&&= 2 \left[\Sigma_{\chi_j}^{>} (x,x') + 
\Sigma_{\chi_j}^{<} (x,x') 
\right] =  2 i {\rm Im}\left[\Sigma_{\chi_j}^{>} (x,x') + 
\Sigma_{\chi_j}^{<} (x,x') 
\right]  \;,
\label{3rd}
\end{eqnarray}

\noindent
where the last equality in (\ref{3rd}) follows from
(\ref{Sigma cond}), from which we see  that
$\Sigma_{\chi_j}^{>} (x,x') +  \Sigma_{\chi_j}^{<} (x,x')$
must be purely imaginary. 

By substituting Eqs. (\ref{2nd}) and (\ref{3rd}) in (\ref{S2}),
due to the result (\ref{3rd}), we are led to an imaginary
contribution to the effective action for the $\chi_j$ fields.
This imaginary term can be appropriately interpreted as coming from
a functional integral over a stochastic field, which then turns the
evolution equation for $\chi_j$ into a stochastic form due
to the presence of a noise term \cite{GR}.
Thus the complete evolution equation for the modes and background fields
includes noise terms. Though their study is particularly difficult,
previous estimates of their effects on the background dynamics during
inflation and reheating \cite{hu} shows that changes in the dynamics
and energy densities are marginal for chaotic inflation kind
of models and within parameters values for coupling constants
($\lambda,g_j,h_k$)
corresponding to the cases of interest in this paper. Due to this
we can neglect the stochastic noise terms appearing
in (\ref{S2}).  On the other hand these noise terms
are important in obtaining the first principles evolution equation
for the fluctuating modes of the inflaton, such as for
studying density fluctuations during inflation;
however in this paper we will not go that far.
Thus by defining the evolution equation for $\chi_j$ modes from
\cite{GR},

\begin{equation}
\frac{\delta S_2[\chi_j^{c},\chi_j^{\Delta}] }
{\chi_j^{\Delta}}\Bigr|_{ \chi_j^{\Delta} = 0 } =0\;,
\label{eom modes}
\end{equation}

\noindent
we are led to the following equation for the $\chi_j$ modes
$f_{\chi_j}({\bf q},t)$ in momentum space

\begin{eqnarray}
\left[ \frac{d^2}{dt^2} + 3 \frac{\dot{a}}{a} 
\frac{\partial}{\partial t} + \frac{\bf{q}^2}{a^2} + M_{\chi_j}^2 (t) \right]
f_{\chi_j}({\bf q},t) +  
\int d t' a^3(t') \Pi_{\chi_j} ({\bf q};t,t')  f_{\chi_j}({\bf q},t') =0 \;,
\label{modes chi}
\end{eqnarray}

\noindent
where $\Pi_{\chi_j}  ({\bf q};t,t')$ is the spatial 
{}Fourier transform of the $\chi_j$
field self-energy term given by Eq. (\ref{2nd}).
An analogous expression for the fluctuation $\phi$ modes also follows,
like Eq. (\ref{modes chi}), with $M_\phi^2 (t) = m_\phi^2 + 
\frac{\lambda}{2} \varphi(t)^2 + \xi R(t)$ and self-energies terms
coming from the dressing due to  $\chi_j$ and $\phi$ loops.
The initial conditions for these field mode differential equations
will be explicitly
stated below for the case of conformal time. Though this is not of
special concern in this work, this is a convenient way
to circumvent known subtle issues 
of renormalization dependence on the initial conditions in gravitational backgrounds
when formulated in comoving time \cite{davis}.

In terms of the general solutions of (\ref{modes chi}),
$f_{1,2} ({\bf q}, t)$ and their
complex conjugate solutions, obtained equivalently from e.g. the
complex conjugate of Eq. (\ref{modes chi}) (note that when
the bath self-energy term entering in (\ref{modes chi}) has
an imaginary part, the equation becomes non-Hermitian), 
we then can write general expressions for the CTP propagator terms, for
both the scalars $\phi$ and $\chi_j$, in
agreement with the continuity conditions expressed in Eq. (\ref{cond}), 
in the general form \cite{semeno,mallik2}

\begin{eqnarray}
&&G^{>} ({\bf q},t,t') = 
f_{1} ({\bf q},t) f_{2}({\bf q},t') \theta(t-t')
+  f^{*}_{1} ({\bf q},t') f^{*}_{2}({\bf q},t)  \theta(t'-t)\;,
\nonumber \\
&& G^{<} ({\bf q},t,t') = 
f_{1}^* ({\bf q},t) f_{2}^*({\bf q},t') \theta(t-t')
+  f_{1} ({\bf q},t') f_{2}({\bf q},t)  \theta(t'-t)\;.
\label{prop-terms}
\end{eqnarray}

The solutions $f_{1,2}({\bf q},t)$ and the appropriated initial conditions
needed to determine them are discussed below.

\section{Solving for the mode functions and real time interacting 
propagators}
\label{solve}

Typically,  equations for the mode functions for an interacting model,
of the general form as given by Eq. (\ref{modes chi}), can be very
difficult to solve analytically, in particular for an 
expanding background. There are, however, a few particular cases, 
like for de Sitter
expansion $H\sim$ constant, so $a(t) = \exp(H t)$, and power law
expansion $a(t) \sim t^n$, where solutions for the mode equation
for free fluctuations are known in exact analytical form
(see e.g. Ref. \cite{davis}). {}For instance
the mode equation for free fluctuations in de Sitter is
(where in this case the scalar of curvature becomes
$R=12 H^2$)

\begin{eqnarray}
\left[ \frac{d^2}{dt^2} + 3 H \frac{d}{dt}+ {\bf q}^2 e^{-2Ht} + m^2 + 
12 \xi H^2 \right]
f_{\rm de \,Sitter}({\bf q},t)  =0 \;,
\label{f de Sitter}
\end{eqnarray}

\noindent
which has known solutions given in terms of Bessel functions
of first and second kind,

\begin{eqnarray}
f_{1,2} ({\bf q},t) \sim J_{\nu} \left(q e^{-Ht}/H \right),\;\;
Y_{\nu} \left(q e^{-Ht}/H \right)\;,
\label{exp_modes}
\end{eqnarray}

\noindent
with $\nu = -i \sqrt{m^2/H^2
+ 12 \xi - 9/4}$. The other case where we can find 
an exact solution for the modes
corresponds to power law expansion, where,
by considering $a(t) = (t/t_0)^n$ and massless (free) fields with
minimal coupling ($\xi=0$), the
solutions are given by \cite{habib}

\begin{equation}
f_{1,2} ({\bf q},t) \sim t^{1/2} H_\mu^{(1)} \left(\frac{q t_0^n t^{1-n}}{n-1}\right),
\;\; t^{1/2} H_\mu^{(2)} \left(\frac{q t_0^n t^{1-n}}{n-1}\right)\;,
\end{equation}

\noindent
with $\mu = (1-3 n)/[2(1-n)]$ and $H_\mu^{(1,2)} (x)=
J_\nu (x) \pm i Y_\nu (x)$ are the Hankel functions.

Alternatively, for deriving an approximate solution for the mode
functions in the interacting case, we can apply a WKB
approximation for equations of the general
form Eq. (\ref{modes chi}) and then check the validity of the
approximation for the parameter and dynamical regime 
of interest to us.
As will be seen below, under
the dynamical conditions we are interested in studying in this paper,
this approximation will suit our purposes. Let
us briefly recall
the WKB approximation and its general validity regime, when applied to
obtaining approximate solutions for field mode equations.
An approximated WKB solution for a mode equation like

\begin{eqnarray}
\left[ \frac{d^2}{dt^2} + \omega^2 ({\bf q},t) \right]
f({\bf q}, t)  =0 \;,
\label{f wkb}
\end{eqnarray}

\noindent
is of the form
$f_{WKB}({\bf q},t)= 1/\left[\omega({\bf q},t)\right]^{1/2}
\exp\left[ \pm i \int^t dt'' \omega({\bf q},t'')\right]$,
which holds under the general
adiabatic condition $\dot{\omega}({\bf q},t) \ll \omega^2({\bf q},t)$.
We must point out that there is no problems in extending this 
approximation to an expanding background and in fact it is a
common approximation taken for instance in the analysis 
of perturbation modes in the adiabatic regime \cite{wkbad}. In that case,
however, massless modes are considered and so the approximation
holds only for large enough physical momenta $q/a\gg H$, corresponding
to wavelengths deep inside the horizon. Here, instead, we
work in the large mass scale regime, e.g. $m_{\chi} \gg H$. In this
regime the WKB approximation is also valid.  This can easily be 
checked by comparing the WKB solution $f_{WKB}$ with the one
obtained from the exact solution for the modes equations, e.g.,
given by (\ref{exp_modes}), $f_{exact}(q,t)$, for free fields
in de Sitter spacetime.
For example, it is useful to examine the ratio
$|f_{WKB}/f_{exact}|$. {}For both solutions the same initial/boundary
conditions are taken (in conformity to the Bunch-Davies vacuum \cite{davis}) 
at $t_0$
(and lets say $t_0=0$), so both results match at the initial time. They
also match in the asymptotic $q \to \infty$ or $H \to 0$ limits, 
as they should, so as to correctly reproduce the Minkowski results.
But they are also found to match very well for masses  $m \gg H$, 
independent of the value of the physical momenta 
(in particular even for $q/a \ll H \ll m$).
{}For example, it can easily be checked that for (in units of $H$) $\xi=0$,
$q/a=0.01 H$ and for $m=10 H$, the overall numerical difference between
the exact and approximated WKB forms for the modes is at most not more
than one percent for an evolution in the first 10 e-folds and this
discrepancy gets smaller for longer evolutions. {}For the typical
parameters we consider in this work (in Sec. VI) and relevant
for the dissipation mechanism discussed in Sec. II, we have for 
example $m_\chi \stackrel{>}{\sim} 10^6 H$, and so the WKB
approximation is expected to be excellent, which is indeed confirmed
by the numerical results to be shown later in this paper.
In addition, note also that large mass scales, $m \gg H$, imply that
curvature effects in the field quantum corrections to be considered for the
background inflaton field are subleading, with the dominant terms being the
Minkowski like corrections.

Proceeding with our derivations, consider then a differential 
equation in the form of Eq. (\ref{modes chi}). 
Instead of working in cosmic time, it is more convenient to work
in conformal time $\tau$, defined by $d \tau = dt/a(t)$, in which case
the metric becomes conformally flat, 

\begin{equation}
ds^2 = a(\tau)^2 \left( d\tau^2 - d {\bf x}^2 \right)\;,
\end{equation}
By also defining a rescaled mode field in conformal time
by

\begin{equation}
\frac{1}{a(\tau)} \bar{f} ({\bf q}, \tau) = f ({\bf q},t) \;,
\label{mode tau}
\end{equation}

\noindent
we can then re-express Eq. (\ref{modes chi}) in the form (generically 
valid for either $\phi$ or $\chi_j$ scalar fluctuations)

\begin{eqnarray}
\frac{d^2}{d \tau^2} \bar{f} ({\bf q}, \tau) + 
\bar{\omega}({\bf q},\tau)^2 \bar{f} ({\bf q}, \tau)
+  \int d \tau' \bar{\Pi} ({\bf q},\tau,\tau')  \bar{f}({\bf q},\tau') =0 \;,
\label{modes tau}
\end{eqnarray}

\noindent
where we have defined

\begin{equation}
\bar{\omega}({\bf q},\tau)^2 = {\bf q}^2+ a(\tau)^2 \left[M^2 +
\left(\xi - \frac{1}{6} \right) R (\tau) \right] \; .
\label{omegabar}
\end{equation}
In (\ref{omegabar}) the conformal symmetry appears in
an explicit form, with $\xi =1/6$ referring to fields conformally coupled to
the curvature, while
$\xi=0$ gives the minimally coupled case. Note also that in conformal time
the scalar curvature becomes 

\begin{equation}
R(\tau) = \frac{6}{a^3} \frac{d^2 a}{d\tau^2}\;.
\end{equation}

\noindent
In Eq. (\ref{modes tau}) we have also
defined the self-energy in conformal time as,

\begin{equation}
\frac{\bar{\Pi} ({\bf q},\tau,\tau')}{a(\tau)^{3/2} a(\tau')^{3/2} } =  
\Pi({\bf q},t,t') \;,
\end{equation}

\noindent
where the self-energy contribution  $\Pi$, coming from the integration over
the bath fields, is given by the space {}Fourier transformed form
for Eq. (\ref{2nd}).  In (\ref{2nd}), $\Pi$ was split into
symmetric and antisymmetric pieces with respect to its argument
as defined in Eq. (\ref{Pis})
Thus based on Eq. (\ref{Pis}), the self-energy term in (\ref{modes tau})
can then be written as
$\bar{\Pi} ({\bf q},\tau,\tau')=\bar{\Pi}_1 ({\bf q},\tau,\tau') +
\bar{\Pi}_2 ({\bf q},\tau,\tau')$.
In addition, by writing the self-energy
term in a diagonal (local) form \cite{ringwald,ian3}

\begin{equation}
\bar{\Pi} ({\bf q},\tau,\tau') = \bar{\Pi} ({\bf q},\tau)
\delta(\tau-\tau') = \left[\bar{\Pi}_1 ({\bf q},\tau) +
\bar{\Pi}_2 ({\bf q},\tau)\right]\delta(\tau-\tau')\;,
\end{equation}

\noindent
and from the properties satisfied by $\Pi_1$ and $\Pi_2$, it results
that $\bar{\Pi}_1 ({\bf q},\tau)$ must be real, while
$\bar{\Pi}_2 ({\bf q},\tau)$ must be purely imaginary.
The real
part of the self-energy contributes to both mass and wave function 
renormalization terms that can be taken into account by a proper redefinition 
of both the field and mass $M$. On the other hand,
the imaginary term of the self-energy is
associated with decaying processes, as discussed previously.
So, we can now  relate  the decay width in terms of the CTP
self-energy terms as

\begin{equation} 
\bar{\Gamma}  =   
-\frac{{\rm Im} \bar{\Pi}} 
{2 \bar{\omega}} = \frac{ \bar{\Sigma}^> - \bar{\Sigma}^< }
{2 \bar{\omega}}\;, 
\label{Gammabar} 
\end{equation} 

\noindent
and Eq. (\ref{modes tau}) can be put in the form

\begin{eqnarray}
\left[ \frac{d^2}{d \tau^2}  + 
\bar{\omega}({\bf q},\tau)^2 
-  2 i  \bar{\omega}({\bf q},\tau) \bar{\Gamma}({\bf q},\tau) \right]
\bar{f}({\bf q},\tau) =0 \;.
\label{modesgamma}
\end{eqnarray}
We can now proceed to obtain a standard WKB solution 
for Eq. (\ref{modesgamma}).  To do this, following the usual WKB
procedure, we assume the solution to have the form $\bar{f}({\bf q},\tau) = 
c \exp[i \gamma({\bf q}, \tau)]$, where $c$ is some constant that can be fixed
by the initial conditions, given by (\ref{mode_cond}) below.
This form of the solution is then substituted into (\ref{modesgamma}) 
to give

\begin{equation}
i \gamma'' - \gamma'^2 + \bar{\omega}^2 -2 i \bar{\omega} \bar{\Gamma} =0\;.
\label{eq gamma}
\end{equation}
Working in the standard WKB approximation, for the 
zeroth order approximation we 
neglect the second derivative term in (\ref{eq gamma}). Then, by taking
$\bar{\Gamma} \ll \bar{\omega}$,
we obtain

\begin{equation}
\gamma_0 \approx \mp \int^\tau_{\tau_0} d\tau' \left(\bar{\omega}-i\bar{\Gamma}\right) \;,
\end{equation}
which is then used back in (\ref{eq gamma}) for the second derivative term to 
determine the next order approximation,

\begin{equation}
\gamma_1 \approx \mp \int^\tau_{\tau_0} d\tau' \left[\bar{\omega}-i\bar{\Gamma}
+ {\cal O} \left( \bar{\omega}'^2/\bar{\omega}^3 \right) \right] 
+i\ln \sqrt{\bar{\omega}} \;.
\end{equation}
The next and following orders in the approximation brings higher
powers and derivatives of $\bar{\omega}'/\bar{\omega}^2 $, which
in the adiabatic regime, $\bar{\omega}'/\bar{\omega}^2 \ll 1$, are
negligible and we are then led to the result

\begin{equation}
\bar{f}_{1,2} ({\bf q},\tau) \approx \frac{c}{\sqrt{\bar{\omega}}} \exp\left[\mp i
\int^\tau_{\tau_0} d\tau' \left(\bar{\omega}-i\bar{\Gamma}\right) \right]\;.
\label{F mode}
\end{equation}
The solutions for the modes of the form (\ref{F mode}) and their complex
conjugate are general
within the adiabatic, or WKB, approximation regime of dynamics.
{}Finally, we completely and uniquely determine the modes by fixing the 
initial conditions at
some initial reference time $\tau_0$, which can be chosen such that
in the limit of $k
\to \infty$ or $H \to 0$ we
reproduce the Minkowski results. These conditions, which correspond to
the ones for the Bunch-Davis vacuum \cite{davis}, 
can be written as

\begin{eqnarray}
&& \bar{f}_{1,2}({\bf q},\tau_0)= \frac{1}{\sqrt{2\bar{\omega}(\tau_0)}} \;,
\nonumber \\
&& \bar{f\;}'_{1,2}({\bf q},\tau_0)= \mp i \sqrt{\bar{\omega}(\tau_0)/2}\;,
\label{mode_cond}
\end{eqnarray}
which already fixes the constant $c$ in (\ref{F mode}) as
$c=1/\sqrt{2}$.

Using the above results in (\ref{prop-terms}) and
after returning to cosmic time $t$, we obtain the result, valid
within the WKB approximation, or adiabatic regime,

\begin{equation}
G^{>(<)} ({\bf q} ,t,t') =  \frac{1}{\left[a(t) a(t')\right]^{3/2} }
\tilde{G}^{>(<)} ({\bf q} ,t,t')\;, 
\end{equation}
where

\begin{eqnarray} 
\tilde{G}^{>}({\bf q} ,t,t') &= & 
\frac{1}{2 [\omega(t)\omega(t')]^{1/2}}
\left\{ 
e^{-i\int_{t'}^t dt'' [\omega(t'')-i\Gamma(t'')]}  
\theta(t-t') +
e^{-i\int_{t'}^t dt'' [\omega(t'')+i\Gamma(t'')]} 
\theta(t'-t) \right\}\;, 
\nonumber \\ 
\tilde{G}^{<}({\bf q} , t,t') & = & \tilde{G}^{>}({\bf q}, t',t) \: , 
\label{G><} 
\end{eqnarray}

\noindent 
where $\Gamma$ is the field decay width in cosmic time,
obtained from (\ref{Gammabar}) and

\begin{equation}
\omega(t) = \sqrt{\frac{{\bf q}^2}{a(t)^2} + M^2(t)}\;,
\end{equation}
with $M^2(t)$, for $\Phi$ particles, given by

\begin{equation}
M^2_\phi(t) = m_\phi^2 +  \frac{\lambda}{2} \varphi(t)^2 + \left(\xi- 
\frac{1}{6} \right) R(t)\;,
\label{Mphi}
\end{equation}
while for $\chi_j$ particles,

\begin{equation}
M^2_{\chi_j}(t) = m_{\chi_j}^2 +  g_j^2 \varphi(t)^2 + \left(\xi- 
\frac{1}{6} \right) R(t)\;.
\label{Mchi}
\end{equation}

The same result (\ref{G><}) could in principle be inferred in an
alternative way by expressing the propagator expressions in terms of a
spectral function, defined by a {}Fourier transform for the difference
between the retarded and advanced dressed propagators, given by Eqs.
(\ref{ret}) and (\ref{adv}), and approximating the spectral function as
a standard Breit-Wigner form with width given by $\Gamma$ and poles
determining the arguments of the exponential in (\ref{F mode}) and its
complex conjugate \cite{GR}. The validity of this approximation in
particular was recently numerically tested and verified in Ref.
\cite{berges1} for a $1+1\;d$ scalar field in Minkowski space-time. In
the Minkowski space-time case, results analogous to Eq. (\ref{G><}) were
explicitly derived in Refs. \cite{GR,ian,BGR,BR}. Indeed, for the case
of no expansion $a(t) = {\rm constant}$, Eq. (\ref{G><}) reproduce the
same analogous expressions as found in the case of Minkowski space-time.

The result (\ref{G><}), from the previous approximations used to derive
the WKB solution (\ref{F mode}), is valid under the requirements

\begin{eqnarray}
&&\Gamma_\phi \ll \omega_\phi\;,
\nonumber \\
&& \Gamma_{\chi_j} \ll \omega_{\chi_j}\;,
\label{Gammaomega}
\end{eqnarray}
and the adiabatic conditions,

\begin{eqnarray}
&& \frac{\bar{\omega}_\phi'}{\bar{\omega_\phi^2}} =
\frac{\dot{a}/a}{\omega_\phi} + \frac{\dot{\omega}_\phi}{\omega_\phi^2}\ll 1 \;,
\nonumber \\
&& 
\frac{\bar{\omega}_{\chi_j}'}{\bar{\omega_{\chi_j}^2}} =
\frac{\dot{a}/a}{\omega_{\chi_j}} + \frac{\dot{\omega}_{\chi_j}}{\omega_{\chi_j}^2}\ll 1 \;,
\label{wkb cond}
\end{eqnarray}

\noindent
where in the second term in the equations (\ref{wkb cond}) we have made the change
back to comoving time and used $\bar{\omega} = a(t) \omega (t)= a
\sqrt{q^2/a^2+M^2}$. The conditions (\ref{Gammaomega}) are
generically valid in perturbation theory. 
The second set
of conditions given by (\ref{wkb cond}) are the usual conditions imposed
in the derivative expansion for the WKB solution (which 
recall was here obtained for
convenience in conformal time).  These conditions 
are valid whenever the adiabatic
conditions for the background field $\varphi(t)$ are satisfied, which is
the case for a slowly moving field. They are also satisfied for those modes
deep inside the horizon, $q \gg a H$, which is useful when expressing
the WKB solution as a large momentum expansion and for explicit
renormalization purposes. {}Finally, the condition (\ref{wkb cond}) is
also found to be satisfied for those modes outside the horizon, $q \ll a
H$, provided the masses $M_\phi$ and $M_{\chi_j}$ are much larger than
the Hubble scale and their time dependence evolves in an adiabatic manner. 
To better see these different regimes of validity of Eq. (\ref{wkb cond}),
we look at the two extreme cases of parameter regimes
of interest in this paper.  For both these cases
$M \gg H$. {}  In the first extreme, the modes are
outside the horizon, $q/a \ll H \ll
M$, in which case Eq. (\ref{wkb cond}) becomes

\begin{eqnarray}
&& 
\frac{\dot{a}/a}{\omega_\phi} + \frac{\dot{\omega}_\phi}{\omega_\phi^2}
\sim \frac{\dot{M}_\phi}{M_\phi^2} \ll 1 \;,
\nonumber \\
&& 
\frac{\dot{a}/a}{\omega_{\chi_j}} + \frac{\dot{\omega}_{\chi_j}}{\omega_{\chi_j}^2}
\sim \frac{\dot{M}_{\chi_j}}{M_{\chi_j}^2} \ll 1 \;,
\label{wkb cond 2}
\end{eqnarray}

\noindent
{}For the other extreme case, the modes are deep inside the horizon,
$q/a \gg M \gg H$, in which case
the conditions Eq. (\ref{wkb cond}) automatically become satisfied, since

\begin{eqnarray}
&& 
\frac{\dot{a}/a}{\omega_\phi} + \frac{\dot{\omega}_\phi}{\omega_\phi^2}
\stackrel{q/a \gg M_\phi \gg H}{\longrightarrow} 0 \;,
\nonumber \\
&& 
\frac{\dot{a}/a}{\omega_{\chi_j}} + \frac{\dot{\omega}_{\chi_j}}{\omega_{\chi_j}^2}
\stackrel{q/a \gg M_{\chi_j} \gg H}{\longrightarrow} 0 \;.
\label{wkb cond 3}
\end{eqnarray}

\noindent
This last case is the weakest condition, since during inflation the
tremendous growth of the scale factor makes the modes rapidly go outside
the horizon, thus going over to the regime of the
first set of constraints  Eq. (\ref{wkb cond 2}).
Condition Eq. (\ref{wkb cond 2}) can be satisfied provided the
background field moves sufficiently slowly, which is the regime we will
be interested in probing in this work. It should also be noted that in
the parameter region of masses $M \gg H$, curvature effects become
subleading and so Minkowski like expressions can apply to leading order.
We therefore expect that in the adiabatic regime
the approximations used to derive Eq. (\ref{G><}) readily hold. This
will be tested numerically later on in Sec. \ref{numerical}.

\section{Deriving the effective equation of motion for the inflaton}
\label{eomderive}

We now turn our attention to the EOM Eq. (\ref{eqphi2}), where we will
work it out in the response theory approximation similar to the
treatment in our recent paper \cite{BR2}. Consider the Lagrangian
density (\ref{Nfields}) in terms of the background (system) field
$\varphi(t)$ and the fluctuation (bath) fields,

\begin{equation}
{\cal L} [ \Phi = \varphi(t)+\phi(x), \chi_j, \bar{\psi}_k, \psi_k, g_{\mu \nu}]
= {\cal L}_\varphi [ \varphi(t), g_{\mu \nu}]
+
{\cal L}_{\rm bath} [\varphi(t),\phi(x), \chi_j, \bar{\psi}_k, \psi_k, g_{\mu \nu}]\;,
\end{equation}

\noindent
where 

\begin{equation}
{\cal L}_\varphi [ \varphi(t), g_{\mu \nu}] = 
a(t)^3 \left\{ \frac{1}{2} \dot{\varphi}(t)^2 - \frac{m_\phi^2}{2}\varphi(t)^2 - 
\frac{\lambda}{4 !} \varphi(t)^4  -\frac{\xi}{2} R \varphi(t)^2
\right\}\;,
\end{equation}

\noindent
is the sector of the Lagrangian independent of the fluctuation bath
fields, while ${\cal L}_{\rm bath}$ denotes the sector of the Lagrangian
that depends on the bath fields. In the following derivation it will be
assumed that the background field $\varphi(t)$ is slowly varying,
something that must be checked for self-consistency. Thus, if we
consider the decomposition of $\varphi(t)$ around some arbitrary time
$t_0$ as $\varphi(t) = \varphi(t_0) + \delta \varphi(t)$, $\delta
\varphi(t)$ can be regarded as a perturbation, for which
a response theory approximation can be used for the derivation of the
field averages in (\ref{eqphi2}). In order to implement the response
theory approximation, we consider the terms in ${\cal L}_{\rm bath}$
that contribute to the derivation of those field averages in the
$\varphi$-EOM Eq. (\ref{eqphi2}) and take $\varphi(t) = \varphi(t_0) +
\delta \varphi(t)$. We denote those terms that depend on $\delta
\varphi(t)$ as ${\cal L}_{\rm int}^{\delta \varphi}$, 

\begin{eqnarray} 
{\cal L}_{\rm int}^{\delta \varphi}  &=&  
a(t)^3 \left\{- 
\frac{\lambda}{4} \left[ 2 \varphi(t_0) \delta \varphi(t) +
\delta\varphi(t)^2 \right] \phi^2 -
\frac{4\lambda}{4 !} 
\delta\varphi(t) \phi^3 
\right. \nonumber \\
&+& \left. \sum_{j=1}^{N_{\chi}} \left[  
- \frac{g_{j}^2}{2} 
\left[2 \varphi(t_0) \delta \varphi(t) +
\delta\varphi(t)^2 \right] \chi_{j}^2  -
g_j^2 
\delta\varphi(t)  \phi \chi_{j}^2
\right]  \right\} 
\:,
\label{Lint} 
\end{eqnarray} 

\noindent
and we treat these terms in ${\cal L}_{\rm int}^{\delta \varphi}$ as
additional (perturbative) interactions.

\subsection{The Response Theory Approximation}

In response theory we express the change in the expectation value of
some operator $\hat{\cal O}(t)$,
$\delta \langle \hat{\cal O}(t) \rangle = 
\langle \hat{\cal O}(t) \rangle_{\rm pert} - \langle \hat{\cal O}(t) \rangle$, 
under the influence of some external
perturbation described by $\hat{H}_{\rm pert}$ which is turned 
on at some time $t_0$, as (for an introductory
account of response theory, see for instance
Ref. \cite{fetter})

\begin{eqnarray}
\delta \langle \hat{\cal O}(t) \rangle 
= i \int_{t_0}^t dt' 
\langle \left[ \hat{H}_{\rm pert}(t'), \hat{\cal O}(t)\right] 
\rangle_{0}\;,
\label{response0}
\end{eqnarray}
where the expectation value on the RHS of (\ref{response0}) is evaluated
in the unperturbed ensemble.
The response function defined by Eq. (\ref{response0}) can be readily 
generalized for the derivation of the field averages. 
Provided that the amplitude $\delta\varphi (t)$ is small
relative to the background field $\varphi(t_0)$,
perturbation theory through the response function can be used to 
deduce the expectation values 
of the fields that enter in the EOM Eq. (\ref{eqphi2}). In this case
the perturbing Hamiltonian $\hat{H}_{\rm pert}$ is obtained from
${\cal L}_{\rm int}^{\delta \varphi}$,
Eq.  (\ref{Lint}). {}From Eqs. (\ref{Lint}) and (\ref{response0})
we can then determine the averages of the bath
fields, for example
$\langle \phi^2(t)\rangle$, as an expansion in $\delta\varphi(t)$, starting
from the time $t_0$ and in an one-loop approximation, as

\begin{eqnarray} 
\langle\phi^2 \rangle \simeq \langle \phi^2 \rangle_0 -
i \int_{t_0}^t 
dt' a(t')^3 \frac{\lambda}{4} \left[2 \varphi(t_0) \delta \varphi(t') +
\delta\varphi(t')^2 \right] 
\langle [\phi^2({\bf x},t),\phi^2({\bf x},t')]
\rangle_0 + {\cal O} (\delta \varphi^3) \;, 
\label{response}
\end{eqnarray}
where $\langle \ldots \rangle_0$ means the correlation function
evaluated for the background field taken at the initial time, $\langle
\ldots \rangle_0 \equiv \langle \ldots \rangle\Bigr|_{\varphi(t_0)}$.
The first term in (\ref{response}) is just the leading order one-loop
tadpole term in the linear response approximation, while the second one is the
one-loop tadpole made up with the interaction term from (\ref{Lint}), $-
a(t)^3 (\lambda/4) \; [2 \varphi(t_0) \delta \varphi(t) +
\delta\varphi(t)^2 ] \phi^2 $, that is used in calculating the leading
order one-loop bubble diagram that gives the two-point function. The
interaction vertex coming from the above term can also be put in the 
more convenient form $-i
a(t)^3 (\lambda/4) \; [\varphi(t)^2 - \varphi(t_0)^2]$, which we
will use in evaluating (\ref{response}).

Using translational invariance we can now write
$\langle [\phi^2({\bf x},t),\phi^2({\bf x},t')]
\rangle$, in Eq. (\ref{response}), in terms of the  causal
two-point
Green's function for the $\phi$ field, $G^{++}_{\phi} (x,x')$, as

\begin{eqnarray}
\lefteqn{\langle [\phi^2({\bf x},t),\phi^2({\bf x},t')]
\rangle = 2 i \: {\rm Im} \langle T \phi^2({\bf x},t)\phi^2({\bf x},t')\rangle}
\nonumber \\
&&~~~~~~~~~~~~= 4 i \:\frac{1}{[a(t)a(t')]^{3}}\int \frac{d^3q}{(2 \pi)^3}
{\rm Im}[\tilde{G}_{\phi}^{++}({\bf q}, t,t')]_{t>t'}^2\;,
\label{response1}
\end{eqnarray}
with $\tilde{G}_{\phi}^{++}({\bf q}, t-t')$ as obtained from Eqs.
(\ref{G of k}) and (\ref{G><}). Eq. (\ref{response}) 
in the response approximation then becomes

\begin{eqnarray}
\langle\phi^2 \rangle \simeq \langle \phi^2 \rangle_0
+
\frac{1}{a(t)^{3}} \int_{t_0}^t 
dt' \lambda \left[\varphi (t')^2 - \varphi(t_0)^2 \right] 
\int \frac{d^3 {\bf q}}{(2 \pi)^3}  
{\rm Im} \left[ \tilde{G}_\phi^{++} ({\bf q},t,t')
\Bigr|_{\varphi(t_0)}  \right]_{t>t'}^2 \;.
\label{lrphi}
\end{eqnarray}

Analogously for the other field averages we find:

\begin{eqnarray}
\langle\chi_j^2 \rangle \simeq \langle \chi_j^2 \rangle_0+
\frac{1}{a(t)^{3}} \int_{t_0}^t 
dt' 2 g_j^2 \left[\varphi (t')^2 - \varphi(t_0)^2 \right] 
\int \frac{d^3 {\bf q}}{(2 \pi)^3}  
{\rm Im} \left[ \tilde{G}_{\chi_j}^{++} ({\bf q},t,t')
\Bigr|_{\varphi(t_0)} \right]_{t>t'}^2
\;,
\label{lrchi}
\end{eqnarray}

\begin{eqnarray}
\langle \phi \chi_j^2 \rangle &\simeq & \langle \phi \chi_j^2 \rangle_0+
\frac{1}{a(t)^{3}} \int_{t_0}^{t} d t' 2 g_j^2 
\varphi (t') 
\frac{1}{[a(t)a(t')]^{3/2}}
\int \frac{d^3 {\bf q}_1}{(2 \pi)^3}  
\frac{d^3 {\bf q}_2}{(2 \pi)^3} \nonumber \\
& \times &
{\rm Im} \left[ \tilde{G}_\phi^{++} ({\bf q}_1,t,t') \tilde{G}_{\chi_j}^{++} 
({\bf q}_2,t,t')   
\tilde{G}_{\chi_j}^{++} ({\bf q}_1+{\bf q}_2,t,t')\right]_{t>t'}
\Bigr|_{\varphi(t_0)} \;,
\label{lrchi2phi}
\end{eqnarray}
and

\begin{eqnarray}
\langle\phi^3 \rangle &\simeq & \langle \phi^3 \rangle_0+
\frac{1}{a(t)^{3}} \int_{t_0}^t 
dt' 2 \lambda \varphi (t') 
\frac{1}{[a(t)a(t')]^{3/2}}
\int \frac{d^3 {\bf q}_1}{(2 \pi)^3}  
\frac{d^3 {\bf q}_2}{(2 \pi)^3} \nonumber \\
&\times & 
{\rm Im} \left[ \tilde{G}_\phi^{++} ({\bf q}_1,t,t') 
\tilde{G}_\phi^{++} ({\bf q}_2,t,t')   
\tilde{G}_\phi^{++} ({\bf q}_1+{\bf q}_2,t,t')\right]_{t>t'}
\Bigr|_{\varphi(t_0)} \;.
\label{lrppp}
\end{eqnarray}

Eqs. (\ref{lrphi}), (\ref{lrchi}), (\ref{lrchi2phi}) and (\ref{lrppp})
are analogous to the ones obtained in \cite{BR} but derived there in
the context of the Schwinger's closed-time-path formalism in Minkowski
space. The leading order terms in the linear response approximation,
$\langle \phi^2 \rangle_0$, $\langle \chi_j^2 \rangle_0$, etc, are
divergent and need appropriate renormalization in expanding
space-time, as e.g. described in Ref. \cite{ringwald2}; below we will
give the explicit expressions for the relevant terms. While
the first two expressions, Eqs. (\ref{lrphi}) and (\ref{lrchi})
correspond, when expressed diagrammatically, to the one-loop tadpoles of
one and two vertices in the $\varphi$-EOM, the last two expressions,
Eqs. (\ref{lrchi2phi}) and (\ref{lrppp}), represent two-loop
contributions to the EOM. In the following, as in Ref. \cite{BR2}, we
will restrict our study of the EOM at one-loop order. This makes our
analysis tractable and simple. Moreover there is no loss in our analysis
of the dissipative dynamics for $\varphi$, since the one-loop terms will
already suffice to demonstrate the possible different dissipative
regimes and the higher order terms only enhance the dissipation effects
obtained in the analysis that follows.

\subsection{The $\varphi$-effective EOM}

We then obtain that the EOM Eq. (\ref{eqphi2}), with bath field averages
evaluated from the response function and at one-loop order, becomes

\begin{eqnarray} 
&& \ddot{\varphi}(t) + 3 \frac{\dot{a}(t)}{a(t)} \dot{\varphi}(t)+
m_\phi^2 \varphi(t) + 
\frac{\lambda}{6} \varphi(t)^3  +\xi R(t) \varphi(t)   \nonumber \\
&&+
\frac{\lambda}{2} \varphi(t) \frac{1}{a(t)^{3}} \int \frac{d^3 q}{(2 \pi)^3} 
\tilde{G}^{++}_\phi({\bf q},t,t)\Bigr|_{\varphi(t_0)}
+ \sum_{j=1}^{N_{\chi}} g_j^2 \varphi (t) \frac{1}{a(t)^{3}}
\int \frac{d^3 q}{(2 \pi)^3} \tilde{G}^{++}_{\chi_j}({\bf q},t,t)
\Bigr|_{\varphi(t_0)}
\nonumber \\
&&+ \frac{\lambda}{2}  \varphi (t) \frac{1}{a(t)^{3}} \int_{t_0}^t dt' \: 
\lambda \left[\varphi(t')^2 - \varphi(t_0)^2 \right]
\int \frac{d^3 {\bf q}}{(2 \pi)^3}  
{\rm Im} \left[ \tilde{G}_\phi^{++} ({\bf q},t,t')\Bigr|_{\varphi(t_0)}
\right]_{t>t'}^2 
\nonumber \\ 
&& + \sum_{j=1}^{N_\chi} g_j^2 \varphi(t)   
\frac{1}{a(t)^{3}}\int_{t_0}^t 
dt' 2 g_j^2 \left[\varphi(t')^2 - \varphi(t_0)^2 \right]
\int \frac{d^3 {\bf q}}{(2 \pi)^3}  
{\rm Im} \left[ \tilde{G}_{\chi_j}^{++} ({\bf q},t,t')
\Bigr|_{\varphi(t_0)} \right]_{t>t'}^2
= 0\;. 
\label{eom1} 
\end{eqnarray} 
 
\noindent
We can now use Eqs. (\ref{G of k}) and (\ref{G><}) and
the equivalent expressions for the $\chi_j$ propagator
in the above equation to obtain

\begin{eqnarray} 
&& \ddot{\varphi}(t) + 3 \frac{\dot{a}(t)}{a(t)} \dot{\varphi}(t)+
m_\phi^2 \varphi(t) + 
\frac{\lambda}{6} \varphi(t)^3  +\xi R(t) \varphi(t)   \nonumber \\
&&+
\frac{\lambda}{2} \varphi(t) \frac{1}{a(t)^{3}} \int \frac{d^3 q}{(2 \pi)^3} 
\frac{1}{2 \omega_\phi(t)}\Bigr|_{\varphi(t_0)}
+ \sum_{j=1}^{N_{\chi}} g_j^2 \varphi (t) \frac{1}{a(t)^{3}}
\int \frac{d^3 q}{(2 \pi)^3} \frac{1}{2 \omega_\chi(t)}
\Bigr|_{\varphi(t_0)}
\nonumber \\
&&- \frac{\lambda^2}{2}  \varphi (t) \int_{t_0}^t dt' \: 
\left[\varphi(t')^2 - \varphi(t_0)^2 \right] D_\phi (t,t') 
- \sum_{j=1}^{N_\chi} 2 g_j^4 \varphi(t)   \int_{t_0}^t 
dt'  \left[\varphi(t')^2 - \varphi(t_0)^2 \right] D_{\chi_j} (t,t')
= 0\;,
\label{eom2} 
\end{eqnarray} 

\noindent
where the kernels $D_\phi (t,t')$ and $D_{\chi_j} (t,t')$ in the above
equation are given, respectively, by

\begin{equation}
D_{\phi} (t,t') = \frac{1}{a(t)^{3}} \int \frac{d^3 {\bf q}}{(2 \pi)^3}  
\sin\left[2\int_{t'}^t dt'' \omega_\phi(t'') \right] \;
\frac{ \exp\left[-2 \int_{t'}^t dt''\Gamma_\phi({\bf q}, t'')\right]  }
{ 4 \omega_\phi(t) \omega_\phi(t')} \Bigr|_{t>t'}\;,
\label{Kphi}
\end{equation}
and

\begin{equation}
D_{\chi_j} (t,t') =\frac{1}{a(t)^{3}} \int \frac{d^3 {\bf q}}{(2 \pi)^3}  
\sin\left[2\int_{t'}^t dt'' \omega_\chi(t'') \right] \;
\frac{ \exp\left[ -2 \int_{t'}^t dt''\Gamma_\chi({\bf q}, t'')\right]  }
{ 4 \omega_\chi(t) \omega_\chi(t')} \Bigr|_{t>t'} \;,
\label{Kchi}
\end{equation}

\noindent
where $\Gamma_\phi$ and $\Gamma_{\chi_j}$ are the decay rates for the
$\Phi$ and $\chi_j$ particles with momentum
$q$, respectively.
These depend explicitly on the
decay channels available for both the $\Phi$ and $\chi_j$ fields,
within the kinematically allowed masses. These we will fix explicitly
below.
Note also that, as a consequence of the linear response approximation,
all the frequencies appearing in the above expressions are
expressed in terms of $\varphi=\varphi(t_0)$,

\begin{eqnarray}
&& \omega_\phi (t) = \left[{\bf q}^2/a(t)^2 + 
m_\phi^2 + \frac{\lambda}{2} \varphi(t_0)^2 + (\xi-1/6) R(t) \right]^{1/2}\;,
\label{omegaphi}
\\
&& \omega_{\chi_j} (t) = \left[{\bf q}^2/a(t)^2 + 
m_{\chi_j}^2 + g_j^2 \varphi(t_0)^2 + (\xi-1/6) R(t) \right]^{1/2} \;.
\label{omegachi}
\end{eqnarray}
In Sec. \ref{Veff-section} we show how the next 
order corrections in the linear response 
approximation (at one-loop order) can be resummed to give back the full
time dependence for $\varphi$ inside the above expressions.

Eq. (\ref{eom2}), with Eqs. (\ref{Kphi}) and (\ref{Kchi}), is our
general expression for the one-loop effective EOM for the background
(inflaton) field $\varphi$. As in the Minkowski space case \cite{BR,BR2}
we expect that the last two, nonlocal terms in Eq. (\ref{eom2}) will
lead to dissipation. This can be made apparent once we integrate them
by parts with respect to $t'$. This way we obtain explicitly first order
(nonlocal) time derivative terms in the background and separate
additional local terms that, when combined with the first two momentum
integral terms appearing in Eq. (\ref{eom2}), will correspond to the
first derivative, $d/d\varphi$, of the one-loop quantum correction to
the effective potential $V_{\rm eff} (\varphi)$ (in the equation
(\ref{eom2}) we have the corrections from both the scalar field $\Phi$
self-coupling and due to its coupling to the $\chi_j$ fields). In the
absence of the nonlocal dissipative terms and the additional couplings
to $\Phi$, this way of obtaining the (field derivative of the) one-loop
effective potential was shown explicitly by Semenoff and Weiss in
\cite{semeno} and its renormalization later studied by Ringwald in
\cite{ringwald2}. 

To make more transparent the interpretation of the different terms that
can be derived from Eq. (\ref{eom2}), let us define here dissipative
kernels $K_\phi(t,t')$ and $K_{\chi_j}(t,t')$ as related to the kernels
Eqs. (\ref{Kphi}) and (\ref{Kchi}), respectively, by 

\begin{equation}
\frac{d K_\phi(t,t')}{d t'} = D_\phi (t,t')\;,
\label{Dphi}
\end{equation}
and

\begin{equation}
\frac{d K_{\chi_j}(t,t')}{d t'} = D_{\chi_j} (t,t')\;,
\label{Dchi}
\end{equation}

\noindent
whose solutions we choose here so that in the limit of flat space
and as we take $t_0 \to -\infty$ the kernels $K_\phi$ and $K_{\chi_j}$ become
the ones obtained in Minkowski space calculations \cite{BR}.
{}From this
we then have the solutions

\begin{equation}
K_\phi(t,t') = \int_{t_0}^{t'} d\tau D_\phi (t,\tau)\;,
\label{sDphi}
\end{equation}
and

\begin{equation}
K_{\chi_j}(t,t') = \int_{t_0}^{t'} d\tau D_{\chi_j} (t,\tau)\;.
\label{sDchi}
\end{equation}

Using Eqs. (\ref{Dphi}) and (\ref{Dchi}) in (\ref{eom2}) we obtain that

\begin{eqnarray} 
&& \ddot{\varphi}(t) + 3 \frac{\dot{a}(t)}{a(t)} \dot{\varphi}(t)+
m_\phi^2 \varphi(t) + 
\frac{\lambda}{6} \varphi(t)^3  +\xi R(t) \varphi(t)   \nonumber \\
&&+
\frac{\lambda}{2} \varphi(t) \frac{1}{a(t)^{3}} \int \frac{d^3 q}{(2 \pi)^3} 
\frac{1}{2 \omega_\phi(t)}\Bigr|_{\varphi(t_0)} 
- \frac{\lambda^2}{2}  \varphi (t)  
\left[\varphi(t)^2 - \varphi(t_0)^2 \right] K_\phi (t,t) 
\nonumber \\
&& + \sum_{j=1}^{N_{\chi}} g_j^2 \varphi (t) \frac{1}{a(t)^{3}}
\int \frac{d^3 q}{(2 \pi)^3} \frac{1}{2 \omega_\chi(t)}
\Bigr|_{\varphi(t_0)}
- \sum_{j=1}^{N_\chi} 2 g_j^4 \varphi(t)   
\left[\varphi(t)^2 - \varphi(t_0)^2 \right] K_{\chi_j} (t,t)
\nonumber \\
&&+ \lambda^2 \varphi (t) \int_{t_0}^t dt' \: 
\varphi(t') \dot{\varphi}(t') K_\phi (t,t') 
+ \sum_{j=1}^{N_\chi} 4 g_j^4 \varphi(t)   \int_{t_0}^t 
dt'  \varphi(t') \dot{\varphi}(t') K_{\chi_j} (t,t')
= 0\;.
\label{eom3} 
\end{eqnarray}

%%%%%%%%%%%%%%%%%%%%%%%%%%

\subsection{The local terms and the effective potential corrections}
\label{Veff-section}

Note that the local terms in the second and third lines in Eq. (\ref{eom3}) can 
be written as a field derivative of the one-loop quantum corrections
from the $\chi_j$ and $\phi$ scalar field fluctuations,
to the effective potential for the $\varphi$ background configuration.
This is easily seen by writing the local terms as

\begin{eqnarray} 
&&\frac{\lambda}{2} \varphi(t) \frac{1}{a(t)^{3}} \int \frac{d^3 q}{(2 \pi)^3} 
\frac{1}{2 \omega_\phi(t)}\Bigr|_{\varphi(t_0)} 
- \frac{\lambda^2}{2}  \varphi (t)  
\left[\varphi(t)^2 - \varphi(t_0)^2 \right] K_\phi (t,t) 
\nonumber \\
&& + \sum_{j=1}^{N_{\chi}} g_j^2 \varphi (t) \frac{1}{a(t)^{3}}
\int \frac{d^3 q}{(2 \pi)^3} \frac{1}{2 \omega_\chi(t)}
\Bigr|_{\varphi(t_0)}
- \sum_{j=1}^{N_\chi} 2 g_j^4 \varphi(t)   
\left[\varphi(t)^2 - \varphi(t_0)^2 \right] K_{\chi_j} (t,t)
\nonumber \\
&& =
\frac{\lambda}{2} \varphi(t) \left\{ \frac{1}{a(t)^{3}} \int \frac{d^3 q}{(2 \pi)^3} 
\frac{1}{2 \omega_\phi}\Bigr|_{\varphi(t_0)} 
- \lambda 
\left[\varphi(t)^2 - \varphi(t_0)^2 \right]  
\frac{1}{a(t)^{3}} \int \frac{d^3 q}{(2 \pi)^3} 
\left[ \frac{1}{8 \omega_\phi^3}\Bigr|_{\varphi(t_0)}  + 
{\cal O}(\Gamma_\phi^2/\omega_\phi^5)
\right]\right\}
\nonumber \\
&& + \sum_{j=1}^{N_{\chi}} g_j^2 \varphi (t) \left\{ \frac{1}{a(t)^{3}}
\int \frac{d^3 q}{(2 \pi)^3} \frac{1}{2 \omega_{\chi_j}}\Bigr|_{\varphi(t_0)} 
-  2 g_j^2  
\left[\varphi(t)^2 - \varphi(t_0)^2 \right] 
\frac{1}{a(t)^{3}} \int \frac{d^3 q}{(2 \pi)^3} 
\left[ \frac{1}{8 \omega_{\chi_j}^3}\Bigr|_{\varphi(t_0)}  + 
{\cal O}(\Gamma_{\chi_j}^2/\omega_{\chi_j}^5)
\right]\right\} \;,
\label{dveff-a} 
\end{eqnarray} 

\noindent
where the ${\cal O}(\Gamma_\phi^2/\omega_\phi^5)$ and 
${\cal O}(\Gamma_{\chi_j}^2/\omega_{\chi_j}^5)$ explicit contributions 
would correspond to higher than one-loop contributions
resulting from the use of the full (resummed) propagators obtained from
Eqs. (\ref{Gphi}) and (\ref{Gchi}). Eq. (\ref{dveff-a}) can now easily be
recognized as originating from the $\delta \varphi$ amplitude expansion of the local
(free) propagators $G_\phi (t,t)$ and $G_{\chi_j}(t,t)$, respectively,

\begin{eqnarray}
G_\phi^{++} (t,t)  &=&
\frac{1}{a(t)^{3}} \int \frac{d^3 q}{(2 \pi)^3} 
\frac{1}{2 \left[{\bf q}^2/a(t)^2 + 
m_\phi^2 + \frac{\lambda}{2} \varphi(t)^2 + (\xi-1/6) R(t) \right]^{1/2}}
\nonumber \\
& = & 
\frac{1}{a(t)^{3}} \int \frac{d^3 q}{(2 \pi)^3} 
\frac{1}{2 \omega_\phi}
- \lambda 
\left[\varphi(t)^2 - \varphi(t_0)^2 \right]  
\frac{1}{a(t)^{3}} \int \frac{d^3 q}{(2 \pi)^3} 
\frac{1}{8 \omega_\phi^3} + {\cal O}(\delta\varphi^3) \;,
\label{Gphitt}
\end{eqnarray}
and

\begin{eqnarray}
G_{\chi_j}^{++} (t,t)  &=&
\frac{1}{a(t)^{3}} \int \frac{d^3 q}{(2 \pi)^3} 
\frac{1}{2 \left[{\bf q}^2/a(t)^2 + 
m_{\chi_j}^2 + g_j^2 \varphi(t)^2 + (\xi-1/6) R(t) \right]^{1/2}}
\nonumber \\
& = & 
\frac{1}{a(t)^{3}}
\int \frac{d^3 q}{(2 \pi)^3} \frac{1}{2 \omega_{\chi_j}}
-  2 g_j^2  
\left[\varphi(t)^2 - \varphi(t_0)^2 \right] 
\frac{1}{a(t)^{3}} \int \frac{d^3 q}{(2 \pi)^3} 
\frac{1}{8 \omega_{\chi_j}^3} 
+ {\cal O}(\delta\varphi^3)
\;.
\label{Gchitt}
\end{eqnarray}

\noindent
This confirms our above statement that these terms arise from
the field derivative of the (unrenormalized) one-loop effective potential 
quantum corrections
coming from the $\Phi$ self-interaction and $\chi_j$ coupling.
Combining the result (\ref{dveff-a}) with the tree level part of
the potential and using (\ref{Gphitt}) and (\ref{Gchitt}), we can then write

\begin{equation}
\frac{d V_{\rm eff} (\varphi,R)}{d \varphi} = m_\phi^2 \varphi(t) + 
\frac{\lambda}{6} \varphi(t)^3  +\xi R(t) \varphi(t) +
\frac{\lambda}{2} \varphi(t) G_{\phi}^{++} (t,t)
+ \sum_{j=1}^{N_{\chi}} g_j^2 \varphi (t) G_{\chi_j}^{++} (t,t)\;.
\label{dveff-b}
\end{equation}

\noindent
The two last terms in Eq. (\ref{dveff-b}) are of course ultraviolet (UV)
divergent as expected and so need to be properly renormalized. This is
done in the usual way by adding to the original Lagrangian, or in the
effective EOM for $\varphi$, the appropriate counterterms of
renormalization, $\delta m_\phi$, $\delta \lambda$ and $\delta \xi$, for
the mass, scalar $\Phi$ self-coupling and the gravitational coupling,
respectively. The details of this renormalization process are discussed
in the Appendix A, where explicit evaluation and renormalization are
done. After renormalization we can just rename the couplings and masses
in the dissipative and quantum corrections as the renormalized ones.
Note also, as evident from Eq. (\ref{dveff-b}) and the explicit results
shown in Appendix A, that the quantum corrections can be kept relatively
under control for perturbative small couplings (and small number of
fields). This is certainly true for the contributions coming from the
$\Phi$ scalar field self-coupling, associated to the inflaton, which is
required to be tiny ($\lambda \lesssim 10^{-13}$) due to the density
perturbations constraints requiring a very flat potential. However, for
the $\chi$ field contributions there are no physical constraints that
require the couplings $g_j$ or the number of fields $N_{\chi}$ to be
sufficiently small. In fact, in Refs. \cite{BR2,BR3} relevant scenarios
of strong dissipation are found for cases of intermediate to large
couplings ${\cal O}(10^{-4}) \lesssim N_\chi g_j \lesssim {\cal O}(1)$.
In these cases we must worry about the large quantum corrections
appearing in (\ref{dveff-b}) and in particular that they will not spoil
the flatness of the potential. We here follow the same procedure adopted
in Refs. \cite{BR2,BR3} to overcome this problem and add to the original
Lagrangian density (\ref{Nfields}) an additional coupling of $\Phi$ to
$N_{\psi_\chi}$ extra fermion fields, $-\sum_{i=1}^{N_{\psi_i}} g'_i \Phi
\bar{\psi}_{i,\chi} \psi_{i,\chi}$, where $\psi_\chi$ are fermion
fields, which are different from the light ones coupled to $\chi$ in
(\ref{Nfields}). {}For appropriately tuned coupling $g' \sim g$ this
modification just mimics supersymmetry, where $\Phi$ couples to both the
boson scalars $\chi$ and their fermion partners, with large
cancellations occurring between the quantum corrections from
the $\chi$ and $\psi_\chi$ fields.
This can be seen explicitly in $V_{\rm
eff}(\varphi)$, when the $\psi_\chi$ fermion coupling to $\Phi$ is
included, which to one-loop order gives

\begin{eqnarray}
V_{\rm eff} (\varphi,R) &=&\frac{m_\phi^2}{2} \varphi^2 + 
\frac{\lambda}{4 !} \varphi^4
+ \frac{\xi}{2} R \varphi^2 \nonumber \\
&+& \frac{1}{2} \int \frac{d^3k_p}{(2\pi)^3}
\left( E_{m_{\phi}} + \sum_{i=1}^{N_{\chi}}  E_{m_{\chi_i}} \right) 
- 2 \int \frac{d^3k_p}{(2\pi)^3}
\sum_{i=1}^{N_{\psi_i}}  E_{m_{\psi_{i,\chi}}},
\label{Veff-unr}
\end{eqnarray}
where $k_p = k/a$ is the physical momentum and

\begin{eqnarray}
&& E_{m_{\phi}}  = 
\sqrt{{\bf k_p}^2 + m_{\phi}^2 + \lambda \varphi^2/2+(\xi-1/6)R},
\nonumber \\
&& E_{m_{\chi_i}}  =  
\sqrt{{\bf k_p}^2 + m_{\chi_i}^2 + g_i^2 \varphi^2+(\xi-1/6)R},
\nonumber \\
&& E_{m_{\psi_{i,\chi}}}  =  \sqrt{{\bf k_p}^2 + (m_{\psi_{i,\chi}} + g'_i \varphi)^2}.
\end{eqnarray}
Thus, with appropriately tuned parameters
$g_i$, $g'_i$ and with zero explicit masses
$m_{\psi_{i,\chi}}=m_{\chi_i}=0$ and $N_{\psi_i}=N_{\chi}/4$\footnote{When the $\chi$
and $\psi_\chi$ couplings to $\Phi$ are treated in a SUSY context, this last restriction
on the number of fields is not important, since $\chi$ are then actually complex
fields while their fermionic partners are Majorana spinors and the two contributions
appear in (\ref{Veff-unr}) with the same number of degrees of freedom.},
the one-loop quantum corrections to $V_{\rm eff}$ cancel to all orders in 
$g_i$, $g^{\chi}_i$ in the nonexpanding case ($R=0$). Even when a vacuum energy is
considered (e.g. in de Sitter where $R=12 H^2$), the combined nonvanishing
contributions from $\chi$ and $\psi_\chi$ can still be made small enough compared
to the three level potential \cite{Vilenkin:sg}.
{}Further discussion about SUSY models is in Sec. VII.

Note also that adding the interaction term $-\sum_{i=1}^{N_{\psi_i}} g'_i
\Phi \bar{\psi}_{i,\chi} \psi_{i,\chi}$ (with the corresponding kinetic
term for the additional fermion species) to the original Lagrangian
density (\ref{Nfields}) will also yield additional contribution to the
EOM for $\varphi$, given by $ \sum_{i=1}^{N_{\psi_i}} g'_i \langle
\bar{\psi}_{i,\chi} \psi_{i,\chi} \rangle$. This term can be worked out
analogous to the scalar case. Aside from the local corrections discussed
above, this term will lead to an additional dissipative kernel in
(\ref{eom3}). As shown explicitly in \cite{BR}, where this term was
derived, it will not be directly proportional to the $\varphi$ field
amplitude, unlike the two nonlocal terms in (\ref{eom3}) coming from the
scalar field quantum corrections that are directly proportional to the
square of $\varphi$. As in the old reheating like scenario discussed in
Sect. \ref{compare}, the fermionic nonlocal term will then be relevant
in the linear regime (or small $\varphi$ amplitude), while the last two
terms in (\ref{eom3}) will contribute mainly in the nonlinear regime of
interest here.  Thus, we can just restrict our following analysis of the
dissipative kernels to the ones given in (\ref{eom3}) and neglect the
contribution to the dynamics coming from the $\psi_{i,\chi}$
interaction, keeping in mind that the addition of any other bath fields
coupled to $\Phi$ will also add to dissipation or nontrivial effects
that can play a role in different dynamical regimes. We will briefly
return to this again in the conclusions in connection to the description
of late time effects in the dynamics of $\varphi$.

\subsection{The effective nonlocal EOM and energy densities system of equations}

With the considerations above, we can now write
the effective equation of motion for the background field
$\varphi$ as

\begin{eqnarray}
\lefteqn{ {\ddot \varphi}(t)  +  3H(t) {\dot \varphi}(t) + 
\frac{dV_{\rm eff}^r(\varphi(t),R(t))}{d\varphi(t)} }
\nonumber \\
&&+ \lambda^2 \varphi (t) \int_{t_0}^t dt' \: 
\varphi(t') \dot{\varphi}(t') K_\phi (t,t') 
+ \sum_{j=1}^{N_\chi} 4 g_j^4 \varphi(t)   \int_{t_0}^t 
dt'  \varphi(t') \dot{\varphi}(t') K_{\chi_j} (t,t')
= 0\;,
\label{eom3a} 
\end{eqnarray} 
where $V_{\rm eff}^r$ stand for the renormalized effective potential
(see Appendix A) and, from (\ref{Kphi}), (\ref{Kchi}), (\ref{sDphi}) and (\ref{sDchi}),

\begin{equation}
K_\phi(t,t') =  \int_{t_0}^{t'} d\tau 
\frac{1}{a(t)^{3}} \int \frac{d^3 {\bf q}}{(2 \pi)^3}  
\sin\left[2\int_{\tau}^t dt'' \omega_\phi(t'') \right] \;
\frac{ \exp\left[-2 \int_{\tau}^t dt''\Gamma_\phi({\bf q}, t'')\right]  }
{ 4 \omega_\phi(t) \omega_\phi(\tau)} \Bigr|_{t>t'}\;,
\label{Kphia}
\end{equation}
and

\begin{equation}
K_{\chi_j}(t,t') =  \int_{t_0}^{t'} d\tau 
\frac{1}{a(t)^{3}} \int \frac{d^3 {\bf q}}{(2 \pi)^3}  
\sin\left[2\int_{\tau}^t dt'' \omega_\chi(t'') \right] \;
\frac{ \exp\left[ -2 \int_{\tau}^t dt''\Gamma_\chi({\bf q}, t'')\right]  }
{ 4 \omega_\chi(t) \omega_\chi(t')} \Bigr|_{t>t'} \;, 
\label{Kchia}
\end{equation}
are the dissipative nonlocal kernels.

The complete evolution of the inflaton field is then determined from
Eq. (\ref{eom3a}) and the Einstein equations for the background
cosmology.  Together, these equations form a complete
set of dynamical equations for both $\varphi$ and the metric. The
Einstein equations for the background cosmology can be
formed in terms of the matter and radiation components as usual by the
equations:

\begin{equation} 
H^2 = \frac{8 \pi G}{3} (\rho_m+ \rho_r) -
\frac{k}{a^2} \;, 
\label{G1} 
\end{equation} 
and

\begin{equation} 
2 \dot{H} + 3 H^2 + \frac{k}{a^2} = - 8 \pi G (p_m +
p_r) \;, 
\label{G2} 
\end{equation}

\noindent 
where $G= 1/m_{\rm
Pl}^2$, with $m_{\rm Pl}$ the Planck mass. The parameter 
$k=0,+1,-1$ for a flat,
closed or open Universe, respectively. In this work we 
only consider the flat case, $k=0$. $\rho_{m (r)}$ and
$p_{m (r)}$ are the energy and pressure densities for matter
(radiation), respectively. We also have the standard relations:

\begin{equation}
\rho_m = \frac{1}{2} \dot{\varphi}^2 
+ V_{\rm eff}^r (\varphi, R)\;,
\label{rhom}
\end{equation}

\begin{equation}
p_m = \frac{1}{2}
\dot{\varphi}^2 
- V_{\rm eff}^r (\varphi,R)
\label{pm}
\end{equation}

\noindent
and $p_{r} = \frac{1}{3} \rho_{r}$.

The matter and radiation energy densities $\rho_m$ and 
$\rho_r$ evolve in time as: 

\begin{eqnarray} 
\dot{\rho}_m &+& 3 H \dot{\varphi}^2  
+ \lambda^2 \varphi (t) \dot{\varphi} (t) \int_{t_0}^t dt' \: 
\varphi(t') \dot{\varphi}(t') K_\phi (t,t') 
\nonumber \\
&+& \sum_{j=1}^{N_\chi} 4 g_j^4 \varphi(t)  \dot{\varphi}(t) \int_{t_0}^t 
dt'  \varphi(t') \dot{\varphi}(t') K_{\chi_j} (t,t')
=0 \;
\label{drhom} 
\end{eqnarray} 

\noindent
and (from the energy conservation law)

\begin{eqnarray} 
\dot{\rho}_r &+& 4 H \rho_r - 
\lambda^2 \varphi (t) \dot{\varphi} (t) \int_{t_0}^t dt' \: 
\varphi(t') \dot{\varphi}(t') K_\phi (t,t') 
\nonumber \\
&-& \sum_{j=1}^{N_\chi} 4 g_j^4 \varphi(t)  \dot{\varphi}(t) \int_{t_0}^t 
dt'  \varphi(t') \dot{\varphi}(t') K_{\chi_j} (t,t')
=0 \;. 
\label{drad} 
\end{eqnarray}

Assuming a flat universe ($k=0$), from Eq. (\ref{G1}), we can consider 

\begin{equation}
\rho_r = \frac{3}{8 \pi G} H^2 - \rho_m =
\frac{3}{8 \pi G} H^2 - \frac{\dot{\varphi}^2}{2} 
- V_{\rm eff}^r (\varphi, R) 
\label{sol rhor}
\end{equation}

\noindent
as the first integral of Eq. (\ref{drad}). Using Eqs. (\ref{G1})-(\ref{pm}), 
we can also express the equation for the acceleration in the following form

\begin{equation}
\frac{\ddot{a}}{a} = \frac{8 \pi G}{3} (\rho_m - \rho_r)-
4 \pi G\dot{\varphi}^2\;.
\label{G3}
\end{equation}

The above equations together with Eq. (\ref{eom3}) form a closed, general
set of (integro-) differential equations for the effective evolution for
the background field $\varphi(t)$ and metric at one-loop and leading
order in the linear response approach.

\subsection{The equation of motion in a local
approximation}

The derived equation of motion for $\varphi$, Eq. (\ref{eom3a}), in
expanding space-time is a considerably more complicated expression
than e.g. the analogous one that would be derived in the Minkowski case.
It is therefore interesting first to see when and whether we can recover
expressions equivalent to the Minkowski space ones in
\cite{BR}. This can be the case for instance if we restrict the dynamics
in the adiabatic regime close to equilibrium, which then requires in
general that the decay rates are larger than the Hubble constant,
$\Gamma \gg H$. Using this and restricting to time intervals $t-t' \sim
1/\Gamma$, where the scale factor consequently changes very little, the
frequency terms in the inner time integrals inside the kernels
expressions Eqs. (\ref{Kphi}) and (\ref{Kchi}) will change very little,
and they can be taken just as constant terms. Under these circumstances
we can then easily see that Eqs. (\ref{Kphi}) and (\ref{Kchi}) can be
approximated to

\begin{equation}
D_{\phi} (t,t') \sim \frac{1}{a(t)^{3}} \int \frac{d^3 {\bf q}}{(2 \pi)^3}  
\sin\left[2 \omega_\phi(t) |t-t'| \right] \;
\frac{ \exp\left[-2 \Gamma_\phi({\bf q}, t) |t-t'| \right]  }
{ 4 \omega_\phi(t)^2} \;,
\label{KphiM}
\end{equation}
and

\begin{equation}
D_{\phi} (t,t') \sim \frac{1}{a(t)^{3}} \int \frac{d^3 {\bf q}}{(2 \pi)^3}  
\sin\left[2 \omega_{\chi_j}(t)  |t-t'|\right] \;
\frac{ \exp\left[ -2 \Gamma_\chi({\bf q}, t) |t-t'|\right]  }
{ 4 \omega_{\chi_j}(t)^2}  \;, 
\label{KchiM}
\end{equation}

\noindent
which are equivalent to the kernels derived in \cite{BR}.
In terms of (\ref{sDphi}), (\ref{sDchi}), (\ref{KphiM}) and (\ref{KchiM}),
the EOM Eq. (\ref{eom3}) now becomes

\begin{eqnarray} 
&& \ddot{\varphi}(t) + 3 H \dot{\varphi}(t)+
\frac{d V_{\rm eff}^r (\varphi,R)}{d \varphi} 
\nonumber \\
&&+ \lambda^2 \varphi (t) \int_{t_0}^t dt' \: 
\varphi(t') \dot{\varphi}(t') \frac{1}{a(t)^{3}}
\int \frac{d^3 q}{(2 \pi)^3} \frac{ \left[\omega_\phi \cos(2 \omega_\phi |t-t'|) + 
\Gamma_\phi  \sin(2 \omega_\phi |t-t'|) 
\right] }{ 8 \omega_\phi^2 \left( \Gamma_\phi^2 + \omega_\phi^2 
\right) } e^{-2 \Gamma_\phi |t-t'|} \nonumber \\ 
&& 
+ \sum_{j=1}^{N_\chi} g_j^4 \varphi(t)   \int_{t_0}^t 
dt'  \varphi(t') \dot{\varphi}(t') \frac{1}{a(t)^{3}}
\int \frac{d^3 q}{(2 \pi)^3}
\frac{ \left[\omega_{\chi_j} \cos(2 \omega_{\chi_j} |t-t'|) + \Gamma_{\chi_j}  
\sin(2 \omega_{\chi_j} 
|t-t'|) \right] }{ 2 \omega_{\chi_j}^2 \left( \Gamma_{\chi_j}^2 + \omega_{\chi_j}^2 
\right) } e^{-2 \Gamma_{\chi_j} |t-t'|} 
= 0\;.
\label{eomfinal} 
\end{eqnarray} 

In this paper we restrict the analysis 
to zero temperature (or in a non-thermalized bath) and use
the same regime of parameters as studied recently in \cite{BR2}, where the
masses (the renormalized and, if relevant, background field dependent
ones) satisfy the condition $M_{\chi_j} > 2 M_{\psi_k} > M_{\phi}$. In
this case $\Gamma_\phi$ vanishes, while $\Gamma_{\chi_j}$ gives the decay
width for the kinematically available decay channel of the scalar
$\chi_j$ fields into the fermion fields $\psi_k,\bar{\psi}_k$. 
To obtain an expression for $\Gamma_{\chi_j}$, first note that
under the condition $M_{\chi_j} \gg H$, 
which is satisfied for the couplings
and values of the inflaton amplitude taken here, the curvature effects
become negligible in the computation of $\Gamma_{\chi_j}$ and, therefore,
it can be well approximated by the Minkowski decay rate at leading order.
Thus, we can write its expression in terms  of the decay rate
in the rest frame, $\Gamma_{\chi_j}(0)$,
and then boost it to give (a similar form for the decay of a scalar
particle into fermions in de Sitter space-time
was also used by Ringwald in Ref. \cite{ringwald})

\begin{eqnarray}
\Gamma_{\chi_j} (t) = \frac{M_{\chi_j}}{\omega_{\chi_j}(t)}
\Gamma_{\chi_j}(0) \;,
\label{Gammachi}
\end{eqnarray}
where $\Gamma_{\chi_j}(0)$ is the standard on-shell rate, as evaluated
in Minkowski space-time \cite{BR},

\begin{equation}
\Gamma_{\chi_j}(0)= \sum_{k=1}^{N_\psi} h_{kj}^2
\frac{M_{\chi_j}}{8\pi}
\left(1-\frac{4 M_{\psi_k}^2}{ M_{\chi_j}^2}\right)^{3/2}\;.
\label{rate}
\end{equation}

\noindent
This result quoted for $\Gamma_{\chi_j}(0)$ based on the Minkowski
space-time result, in the regime of field mass $M \gg H$, is corroborated
by the derivation of a decay rate expression in de Sitter space-time shown 
in Ref. \cite{boya1}, where it was shown that the decay rate behaves similar to
the Minkowski one, but in a thermal bath at the Hawking temperature.
However since in our calculation $M_{\chi} \gg H$, this modification has
negligible effect. There is also a simple way of understanding the
result (\ref{Gammachi}) in the case of an expanding space-time within
the regime of parameters we are examining. Recall in conformal time
(with conformal rescaled fields) our original model is no different from
the one of a Minkowski space-time, except for the proper rescalings of
dispersion relations and masses, $\omega=\sqrt{q^2/a^2 + M^2} \to
\bar{\omega} = \sqrt{q^2 + \bar{M}^2}$, where $\bar{M} = a M$. Thus a
rate evaluated in a conformally invariant theory, in conformal
variables, is identical to that in flat space-time. {}For instance,
consider the change of number of particles given in its simplest form,
in conformal time, as

\begin{equation}
\frac{d n}{d \tau} \sim  \bar{\Gamma} n\;,
\end{equation}
which in terms of physical time becomes

\begin{equation}
\frac{d n}{d t} \sim  \frac{d \tau}{dt} \bar{\Gamma} n = \frac{\bar{\Gamma}}{a} n\;.
\end{equation}
The above expression explicitly displays the (conformal) 
rate as suppressed by the scale factor.
However take the case of, e.g., fermion production as given before, but
in conformal rescaled quantities. Since the rate in conformal variables is
identical to that of flat space-time, we have that 

\begin{equation}
\frac{\bar{\Gamma}}{a} = \frac{1}{a} \frac{\bar{M}_{\chi_j}}{\bar{\omega}_{\chi_j}}
\sum_{k=1}^{N_\psi} h_{kj}^2
\frac{\bar{M}_{\chi_j}}{8\pi}
\left(1-\frac{4 \bar{M}_{\psi_k}^2}{ \bar{M}_{\chi_j}^2}\right)^{3/2}
= \frac{M_{\chi_j}}{\omega_{\chi_j}} 
\sum_{k=1}^{N_\psi} h_{kj}^2
\frac{M_{\chi_j}}{8\pi}
\left(1-\frac{4 M_{\psi_k}^2}{ M_{\chi_j}^2}\right)^{3/2}\;,
\end{equation}
which then reproduces the above stated result Eq. (\ref{Gammachi}). 

We now substitute $\Gamma_\phi=0$ and $\Gamma_\chi$ 
given by (\ref{Gammachi})
in (\ref{eomfinal}) and consider the parameter regime
relevant to our analysis here, where the couplings are 
$\lambda \sim {\cal O}(10^{-13})$ and $g,g',h \gtrsim {\cal O}(10^{-2})$.
In this regime,
we can generically drop the first nonlocal term in (\ref{eomfinal}),
which comes from the $\Phi$ scalar self-coupling,
since this term is much smaller in magnitude than the dissipative term
due to the $\chi$ corrections. Equivalently stated, 
since for the above parameter
values the contribution to the dynamics of the background field
$\varphi$ coming from the $\phi$ quantum modes 
are neglegible compared to those due
to the $\chi_j$ ones, we could as well consider from the beginning 
the original inflaton field $\Phi$ in Eq. (\ref{Nfields}) as 
simply a classical
(homogeneous) field $\Phi \equiv \varphi(t)$ in interaction with the 
remaining (quantum) fields in (\ref{Nfields}). This was, for instance,
the approach taken in \cite{BR3}.
Adopting this approach, the only effect on the above calculations
would be to drop all quantum inflaton self-interactions and thus keep only the
nonlocal $\chi$ term in  Eqs. (\ref{eom3a}) and (\ref{eomfinal}).
This point is important to note, since
for the parameter regime to be considered here, we will see that
while $M_\chi \gg H$ the same does not hold true for the inflaton mass,
$M_\phi \sim H$, and so applying
the WKB approximation for the $\phi$ quantum modes could become questionable.
However the smallness of the $\phi$ self-coupling allows us to
completely ignore the effects from the $\phi$ quantum fluctuations.
Thus, in what follows
the quantum effects leading to the $\varphi$-effective EOM will 
only arise
from the terms associated with the $\chi_j$ 
dynamical quantum corrections in Eqs. (\ref{eom3a}) and
(\ref{eomfinal}).
  
{}Further
approximations can be applied to (\ref{eomfinal}) in the
dynamical regime for which the motion of $\varphi$ is slow. In this
case
an adiabatic-Markovian approximation can be applied 
to the nonlocal $\chi$-
contributions.  This converts 
(\ref{eomfinal}) to one that is completely local in time, albeit with time
derivative terms. The details of this approximation
for Minkowski space-time can be found in \cite{BR}. Its extension
to an expanding FRW background follows analogous lines.  
The Markovian approximation amounts to substituting $t' \rightarrow t$ in
the arguments of the $\varphi$-fields in the second nonlocal term 
in Eq. (\ref{eomfinal}).
The adiabatic approximation then requires self-consistently
that all macroscopic motion is
slow on the scale of microscopic motion, thus
${\dot \varphi}/{\varphi},H < \Gamma_{\chi}$.
Moreover when $H < M_{\chi}$,
the kernel $K_\chi(t,t')$ is well approximated by the
nonexpanding limit $H \rightarrow 0$. 
The validity of all these approximations were examined
in \cite{BR3} and they also will be examined in more detail
in Sec. \ref{numerical}.
The result of these approximations is that, after 
trivially integrating over the momentum integral in the last term
in (\ref{eomfinal}), the effective EOM Eq. (\ref{eomfinal}) 
becomes \cite{BR3}

\begin{equation}
{\ddot \varphi} + [3H+\Upsilon(\varphi)] {\dot \varphi} +
\frac{dV_{\rm eff}^r (\varphi,R)}{d\varphi} = 0 .
\label{amapprox}
\end{equation}
By setting the couplings $g_j =g'_j=g, h_{kj}=h\sim g \gg \lambda$,
the mass
$M_{\chi} \simeq g\varphi \gg m_{\psi_k}$ and $\Gamma_{\chi} \simeq N_\psi h^2
M_{\chi}^2/[8\pi \omega_{\chi}]$, it leads to the friction coefficient
$\Upsilon(\varphi)$ in (\ref{amapprox})

\begin{equation}
\Upsilon(\varphi) = N_\chi \frac{\sqrt{2} g^4 \alpha_\chi \varphi^2 }
{64\pi M_\chi \sqrt{1 + \alpha^2_{\chi}}
\sqrt{\sqrt{1 + \alpha^2_{\chi}}+1}},
\label{upsilon}
\end{equation}
where $\alpha_\chi \equiv N_\psi h^2 /(8 \pi )$.

In terms of the approximations used to derive (\ref{amapprox}) the
equations (\ref{drhom}) and (\ref{drad}) also simplify.
In particular, from (\ref{drad}), 
we obtain that

\begin{equation}
\rho_r \sim \Upsilon(\varphi) \frac{\dot{\varphi}^2}{4H}\;.
\label{rhor}
\end{equation}

\section{Numerical Analysis of EOM}
\label{numerical}

This section examines numerical results obtained from the basic
equations that have been derived in the previous sections. In
particular, we follow the considerations taken in the previous section
as regarding the contribution of the nonlocal terms in the
$\varphi$-effective EOM, dropping the neglegible $\phi$ quantum
corrections and keeping only the leading correction to the dynamics,
given by the $\chi_j$ nonlocal term. The behavior of the dissipative
kernel coming from the $\chi_j$ nonlocal term and solutions to the
$\varphi$-effective EOM in the various approximations are then
determined. Also, radiation production is studied in the different
approximations. {}Finally the adiabatic conditions underlying the
self-consistency of the basic equations in this paper are examined.

\subsection{The Dissipation Kernel}

Our analysis starts with the dissipative kernel. {}From $D_{\chi}(t,t')$
given in Eq. (\ref{Kchi}) and specializing our computations to the case
of a de Sitter metric as appropriate for describing the inflationary
phase, we can directly perform the time integrals appearing in
$D_{\chi}(t,t')$ and then use the resulting expression in the dissipative kernel
$K_{\chi}(t,t')$, Eq. (\ref{sDchi}), to obtain 

\begin{equation}
K_{\chi}(t,t') \approx \int \frac{d^3q}{(2\pi)^3}
\frac{\omega_{\chi}({\bf q} a(t),t)\cos[2W_\chi({\bf q} a(t),t,t')]
+\Gamma_\chi ({\bf q}a(t),t) \sin[2W_\chi({\bf q} a(t),t,t')]}
{\omega_\chi ({\bf q}a(t),t)\omega_\chi ({\bf q}a(t),t')
\left[\omega^2_\chi({\bf q}a(t),t) + \Gamma^2_{\chi}({\bf q}a(t),t)
\right]}
E_\chi ({\bf q} a(t),t,t')
\label{kapprox}
\end{equation}
where 

\begin{eqnarray}
W_{\chi}({\bf q},t_1,t_2) \equiv \int_{t_1}^{t_2} dt'' 
\omega_{\chi}({\bf q}, t'') & = &
-\frac{1}{H} \left\{ \omega_\chi({\bf q}, t) - \omega_\chi(t')  + 
\frac{1}{2} \sqrt{m_\chi^{2} + 2(6 \xi -1) H^2} \right.
\nonumber \\
& \times & \left.
\ln \left[  \left( \frac{ \omega_\chi(t) - \sqrt{m_{\chi}^{2} + 2(6 \xi -1) H^2} }
{ \omega_\chi(t') - \sqrt{m_\chi^{2} + 2(6 \xi -1) H^2} }  \right)
\left(
\frac{ \omega_\chi(t') + \sqrt{m_\chi^{2} + 2(6 \xi -1) H^2} }
{ \omega_\chi(t) + \sqrt{m_\chi^{2} + 2(6 \xi -1) H^2}  } \right)
\right]  \right\} \;.
\end{eqnarray}
and

\begin{eqnarray}
E_{\chi}({\bf q},t_1,t_2) & \equiv &
\exp\left[ -2 m_{\chi} 
\Gamma_{\chi_j}(0)
\int_{t_1}^{t_2} dt''1/\omega_{\chi}(t'') \right]
\nonumber \\
& = & \left\{\frac{\left[\omega_{\chi}(t) + \sqrt{m_{\chi}^2+ 
2 (6 \xi -1)H^2} \right] 
\left[ \omega_{\chi}(t') - \sqrt{m_{\chi}^2+ 
2 (6 \xi -1)H^2} \right] }
{\left[\omega_{\chi}(t) - \sqrt{m_{\chi}^2+ 
2 (6 \xi -1)H^2} \right] 
\left[\omega_{\chi}(t')+ \sqrt{m_{\chi}^2+ 
2 (6 \xi -1)H^2}  \right] }
\right\}^{- \Gamma_{\chi}(0)/H}\;.
\end{eqnarray}
Note in arriving at Eq. (\ref{kapprox}),
the factor of $1/a(t)^3$ in Eq. (\ref{Kchi})  has been absorbed by a
change of variable on the momentum integration. 
The solution for $K_{\chi}$ in Eq. (\ref{kapprox}) is valid up to errors
of ${\cal O}({\dot \omega}/{\omega}^2)$.  It is useful to examine the behavior
not only of $K_{\chi}(t,t')$ but also the integrated kernel  

\begin{equation}
I_{\chi}(t) \equiv \int_0^{t} dt' K_{\chi}(t',0).
\label{intker}
\end{equation}

\begin{figure}[ht]
\vspace{1cm}
\epsfysize=14cm
{\centerline{\epsfbox{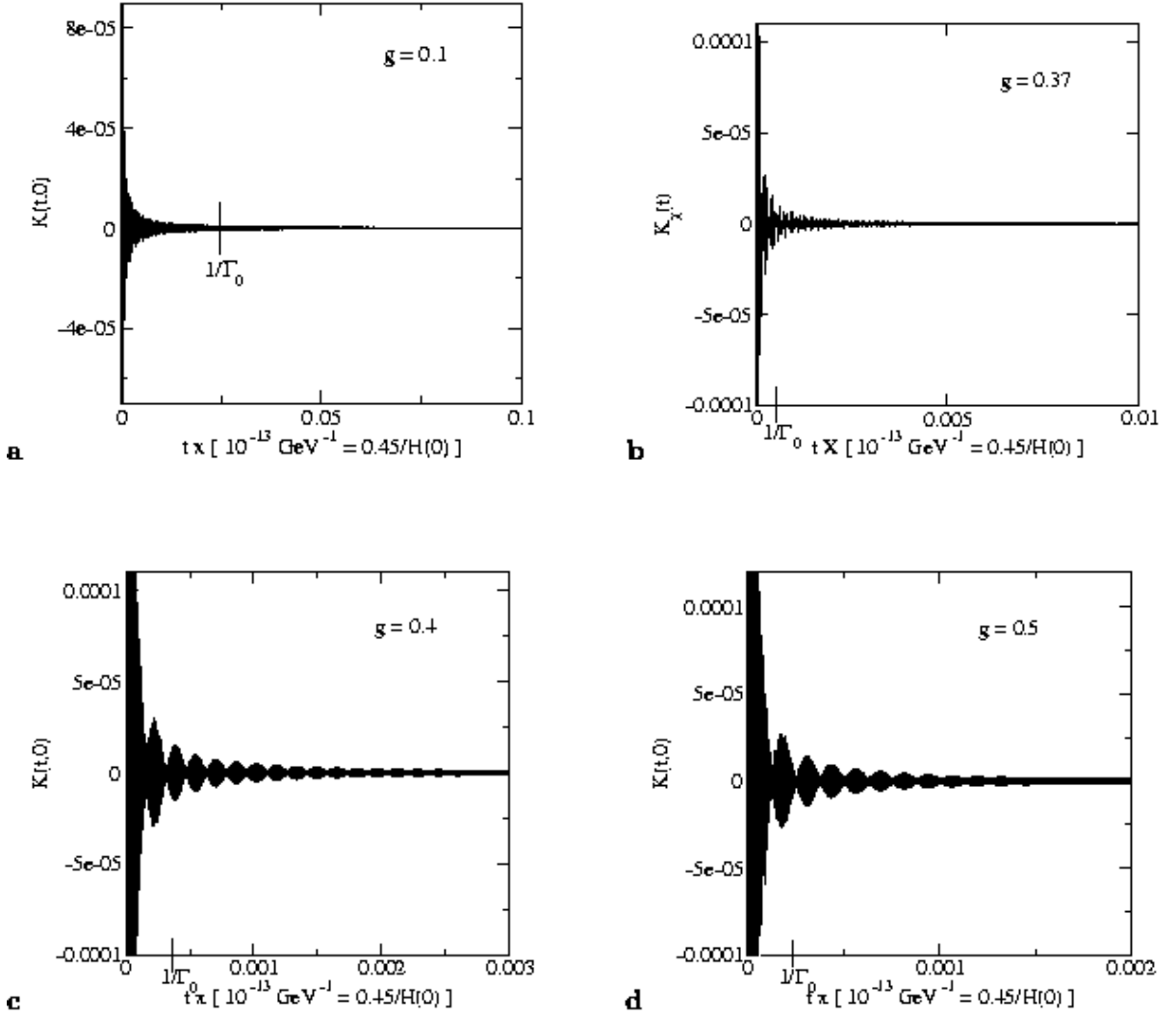}}}
\caption{The kernel $K_\chi(t,0)$ of Eq. (\protect\ref{kapprox}) for various
interaction couplings $g$ and $h=g$, with $m_{\chi}=g \varphi(0)$,
$\varphi(0)= m_{\rm Pl}$, $\lambda=10^{-13}$,
and $\xi=0$.}
\label{figk}
\end{figure}

\begin{figure}[ht]
\vspace{1cm}
\epsfysize=14cm
{\centerline{\epsfbox{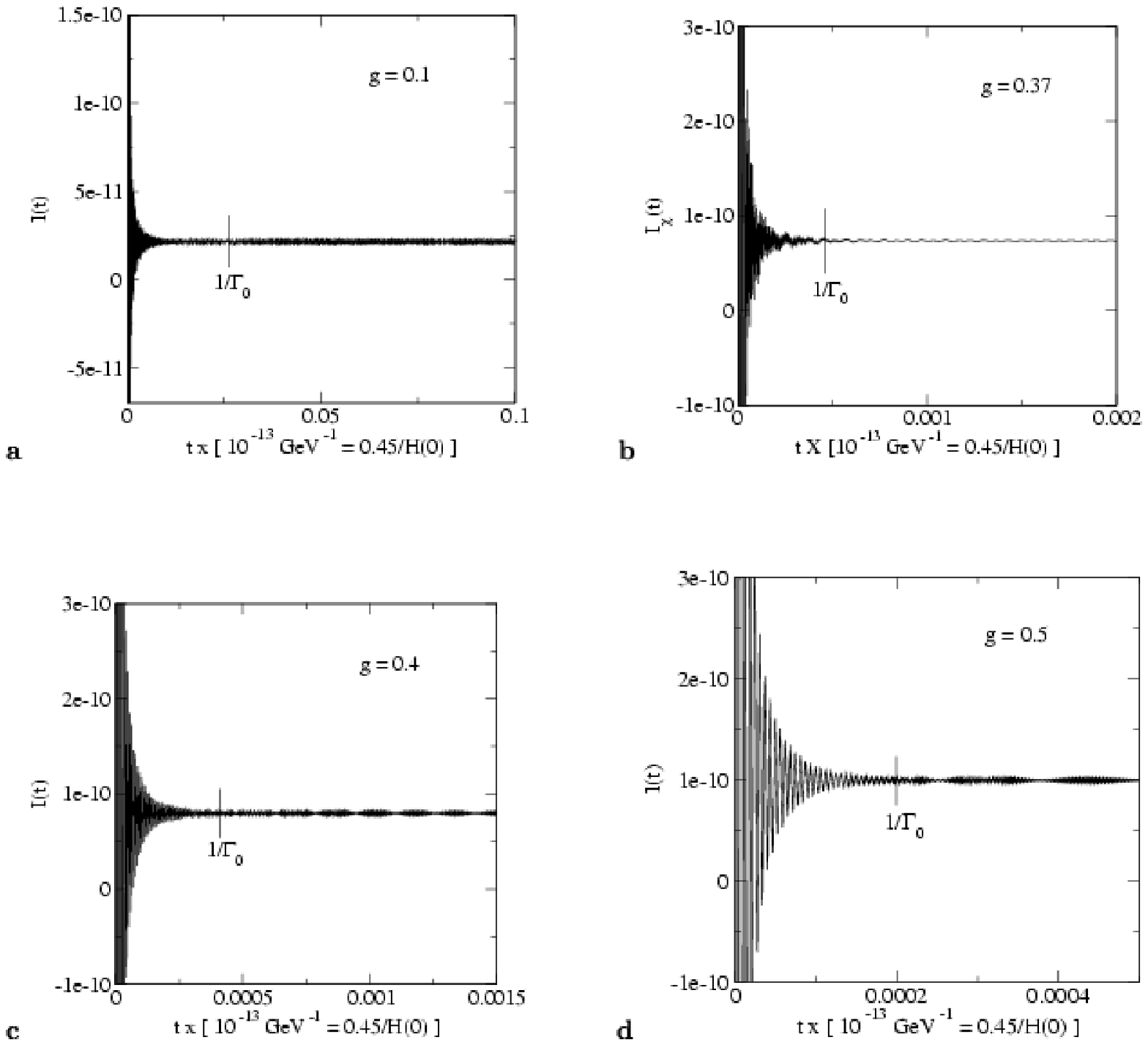}}}
\caption{The integrated kernel $I_\chi(t,0)$ of Eq. (\protect\ref{intker}) 
for various
interaction couplings $g$ and $h=g$, with $m_{\chi}=g \varphi(0)$,
$\varphi(0)= m_{\rm Pl}$, $\lambda=10^{-13}$,
and $\xi=0$.}
\label{figki}
\end{figure}

In Figs. {\ref{figk} and \ref{figki}, 
$K_{\chi}(t,0)$ and $I_{\chi}(t)$ respectively are plotted
for the cases $g=0.1,0.37,0.4$ and $0.5$ in frames
a-d respectively.  In all the graphs the time interval $1/\Gamma_0(g)$
has been indicated, where $\Gamma_0(g)\equiv \Gamma_{\chi}(0)$ as defined
in Eq. (\ref{rate}) is the $\chi$ decay width at zero momentum
${\bf q} = 0$.  The kernel
$K_{\chi} (t,0)$ is seen to oscillate about zero
with an overall enveloping amplitude that
decays in a time interval $\sim 1/\Gamma_{\chi}$.  The graphs
of the integrated kernels $I_{\chi}(t)$,
{}Fig. \ref{figki}, show that there is an
overall skewness, and within the time
interval of order $1/\Gamma_{\chi}$,  the integrated
kernels converge to almost constant values.  Thus, although
the kernel does not have a simple Gaussian or exponential
decay behavior, the rapid oscillatory behavior that it does
have effectively causes it to
retain memory only over a time interval of order
$1/\Gamma_\chi$.  It is also interesting to compare the kernel
for the nonexpanding versus expanding cases, which is shown
in Figs. \ref{figk37comp} and \ref{figki37comp} 
for $K_{\chi}(t,0)$ and $I_{\chi}(t)$ respectively
at coupling $g=0.37$.  The graphs show very little difference
between the two cases, which was expected since 
$m_{\chi} \gg H$, and here is explicitly confirmed. 
{}For the parameters used in the figures we have that
$m_\chi \gtrsim 10^6 H$.  What differences there are between 
the nonexpanding and expanding space-time kernels become increasingly
pronounced as $t$ increases.  This also is expected, since for very
early times $t \ll 1/H$, the effect of expansion should be negligible.
Note also in comparing Fig. \ref{figki} with Fig. \ref{figki37comp}
the y-axis is much more refined in the latter to help facilitate
the desired comparison.  However because of this in 
Fig. \ref{figki37comp} the rapid decay of the integrated kernel
below $t < 1/\Gamma_0$ can not be seen.

\begin{figure}[ht]
\vspace{1cm}
\epsfysize=7cm
{\centerline{\epsfbox{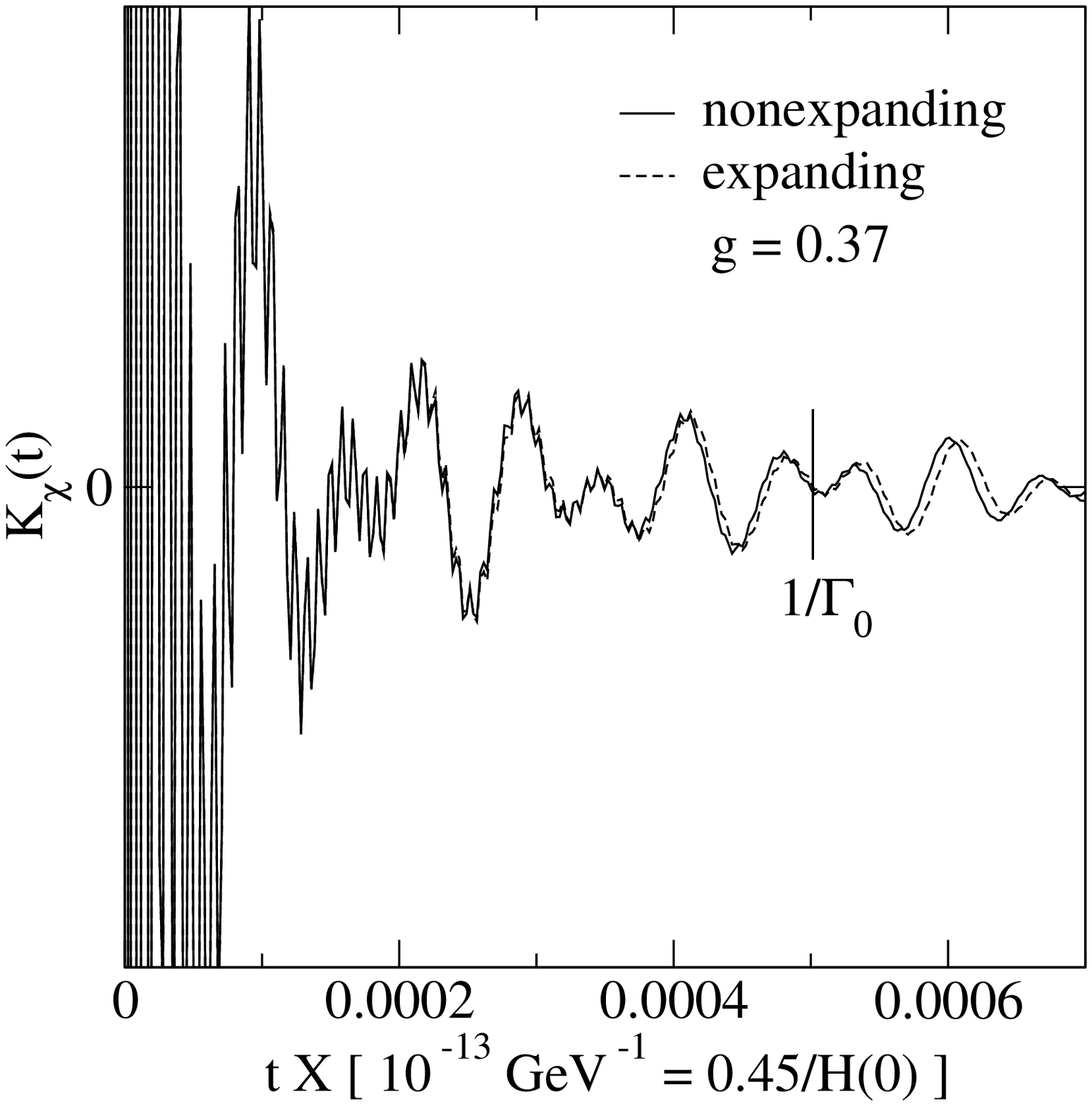}}}
\caption{The kernel compared between the expanding 
Eq. (\protect\ref{kapprox})
and nonexpanding cases for $g=h=0.37$,
with $m_{\chi}=g \varphi(0)$,
$\varphi(0)= m_{\rm Pl}$, $\lambda=10^{-13}$,
and $\xi=0$.}
\label{figk37comp}
\end{figure}

\begin{figure}[ht]
\vspace{1cm}
\epsfysize=7cm
{\centerline{\epsfbox{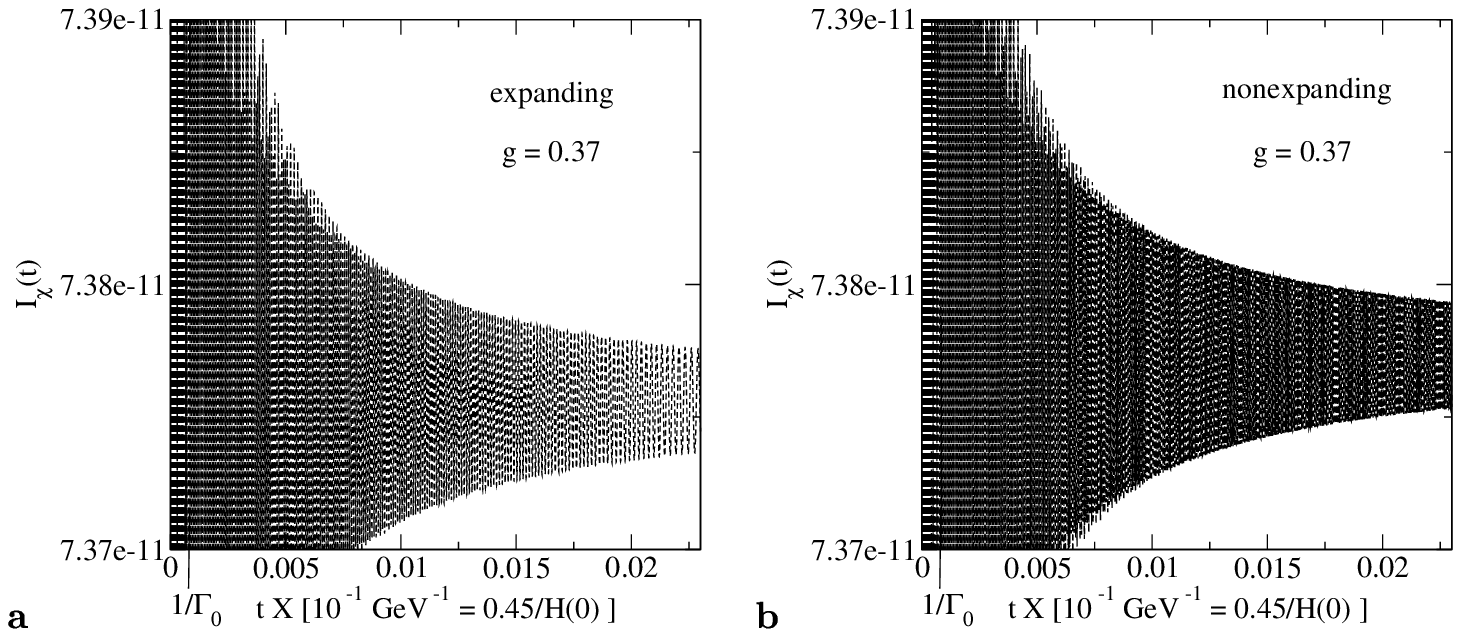}}}
\caption{The integrated kernel of {}Fig. \protect\ref{figk37comp}
compared between the a. expanding 
and b. nonexpanding cases for $g=h=0.37$,
with $m_{\chi}=g \varphi(0)$,
$\varphi(0)= m_{\rm Pl}$, $\lambda=10^{-13}$,
and $\xi=0$.}
\label{figki37comp}
\end{figure}

{}From the graphs of $I_{\chi}(t)$, the origin and validity of the local
approximation Eq. (\ref{amapprox}) of the $\varphi$-evolution equation
Eq. (\ref{eom3a}) can be understood. {}For this, first note that
irrespective of the effects that dissipative damping have on slowing the
evolution of $\varphi$, a minimal damping always arises from the
$3H{\dot \varphi}$ term combined with the flatness of the potential,
which in particular imply that within a time interval $\sim 1/H$,
$\varphi$ and $\dot \varphi$ do not change significantly. In particular
in integrating over the temporally nonlocal term in the $\varphi$-EOM
Eq. (\ref{eom3a}) over a time interval of order $1/H$, $\varphi$ and
$\dot \varphi$ can be treated as constant
and so taken out of the time integration. This leaves integration over
only $K_{\chi}(t,0)$ and as shown in Fig. \ref{figki}, within a time
interval $\sim 1/\Gamma_{\chi}$, this integral rapidly converges to an
almost constant value, which is precisely $\Upsilon/\varphi^2$ of Eq.
(\ref{upsilon}). Since $\Gamma_{\chi} \gg H$ in all the cases in Fig.
\ref{figki} it also means $I_{\chi}(t)$ converges within a time $t \ll
1/H$.

\begin{figure}[ht]
\vspace{1cm}
\epsfysize=8cm
{\centerline{\epsfbox{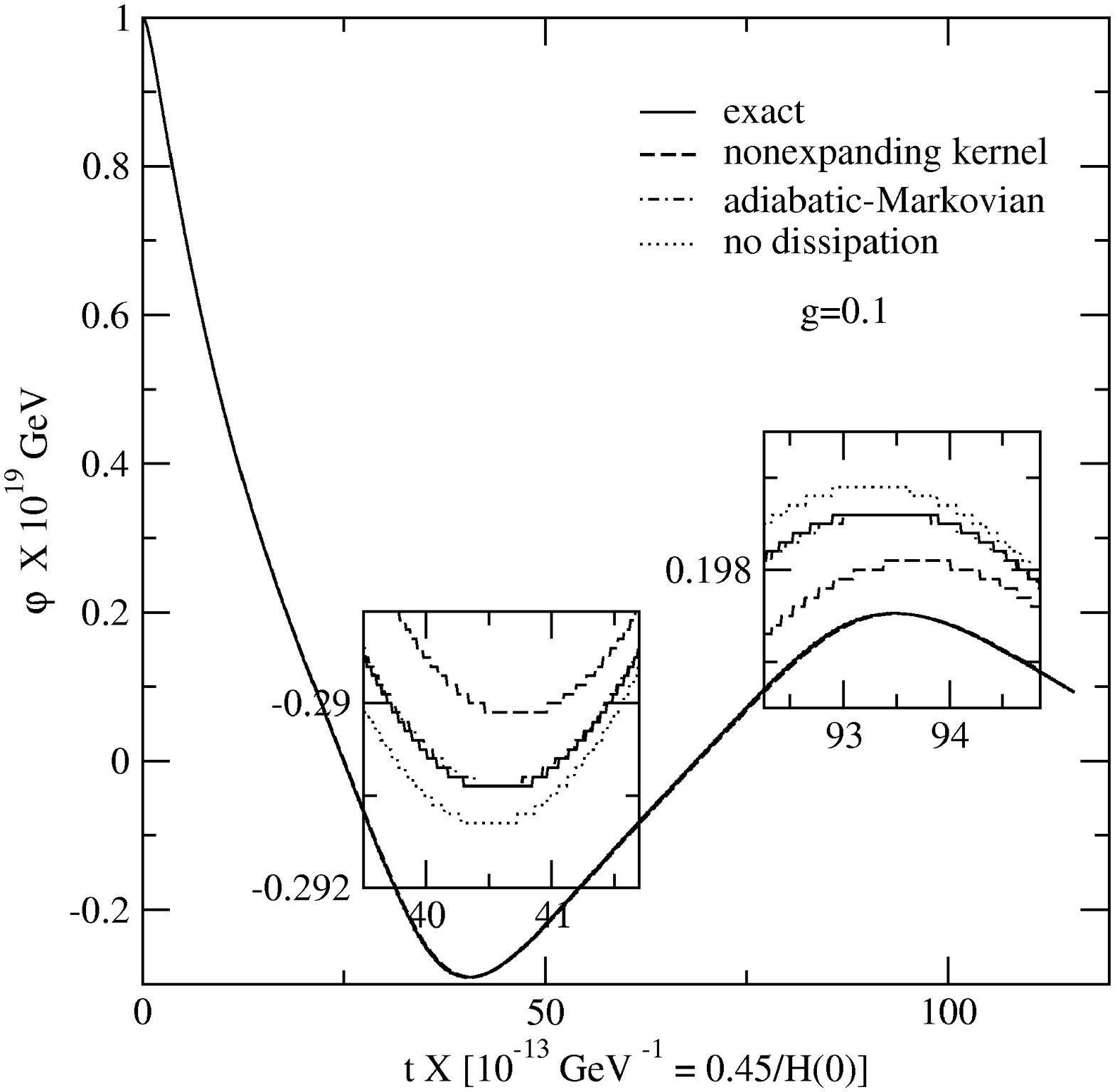}}}
\caption{Evolution of $\varphi(t)$ for $g=h=0.1$, $\lambda=10^{-13}$,
$\xi=0$,
$\varphi(0) = m_{\rm Pl}$, and ${\dot \varphi}(0)=0$.}
\label{figd01}
\end{figure}

\begin{figure}[ht]
\vspace{1cm}
\epsfysize=6.0cm
{\centerline{\epsfbox{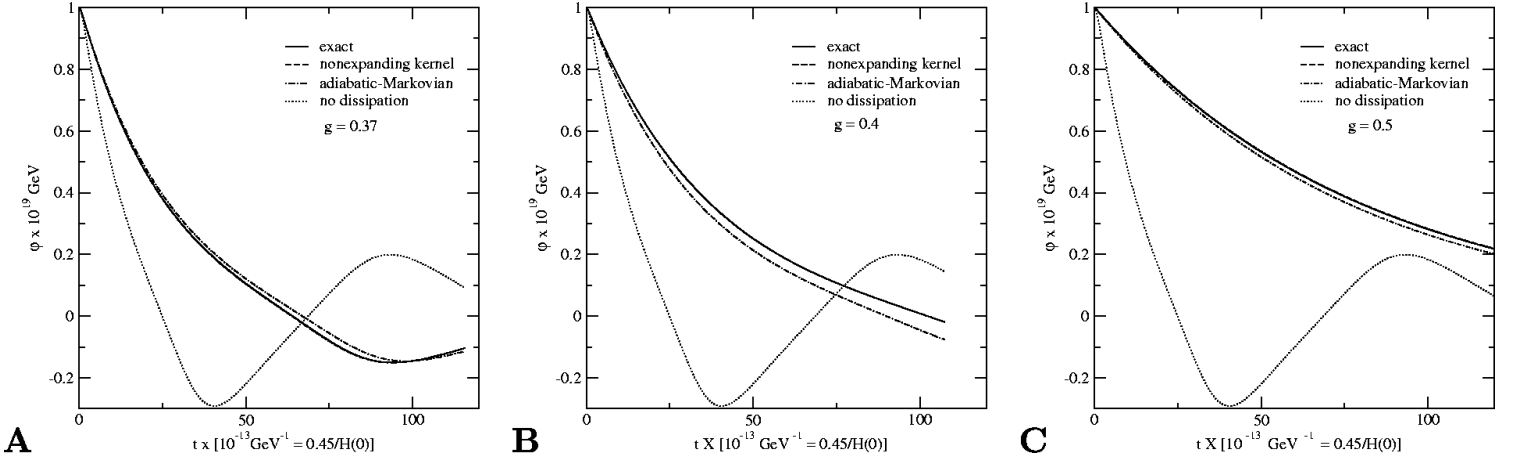}}}
\caption{Evolution of $\varphi(t)$ for various
interaction couplings $g$ and $h=g$, with  $\lambda=10^{-13}$,
$\xi=0$,
$\varphi(0) = m_{\rm Pl}$, and ${\dot \varphi}(0)=0$.}
\label{figvp2}
\end{figure}

\begin{figure}[ht]
\vspace{1cm}
\epsfysize=14.0cm
{\centerline{\epsfbox{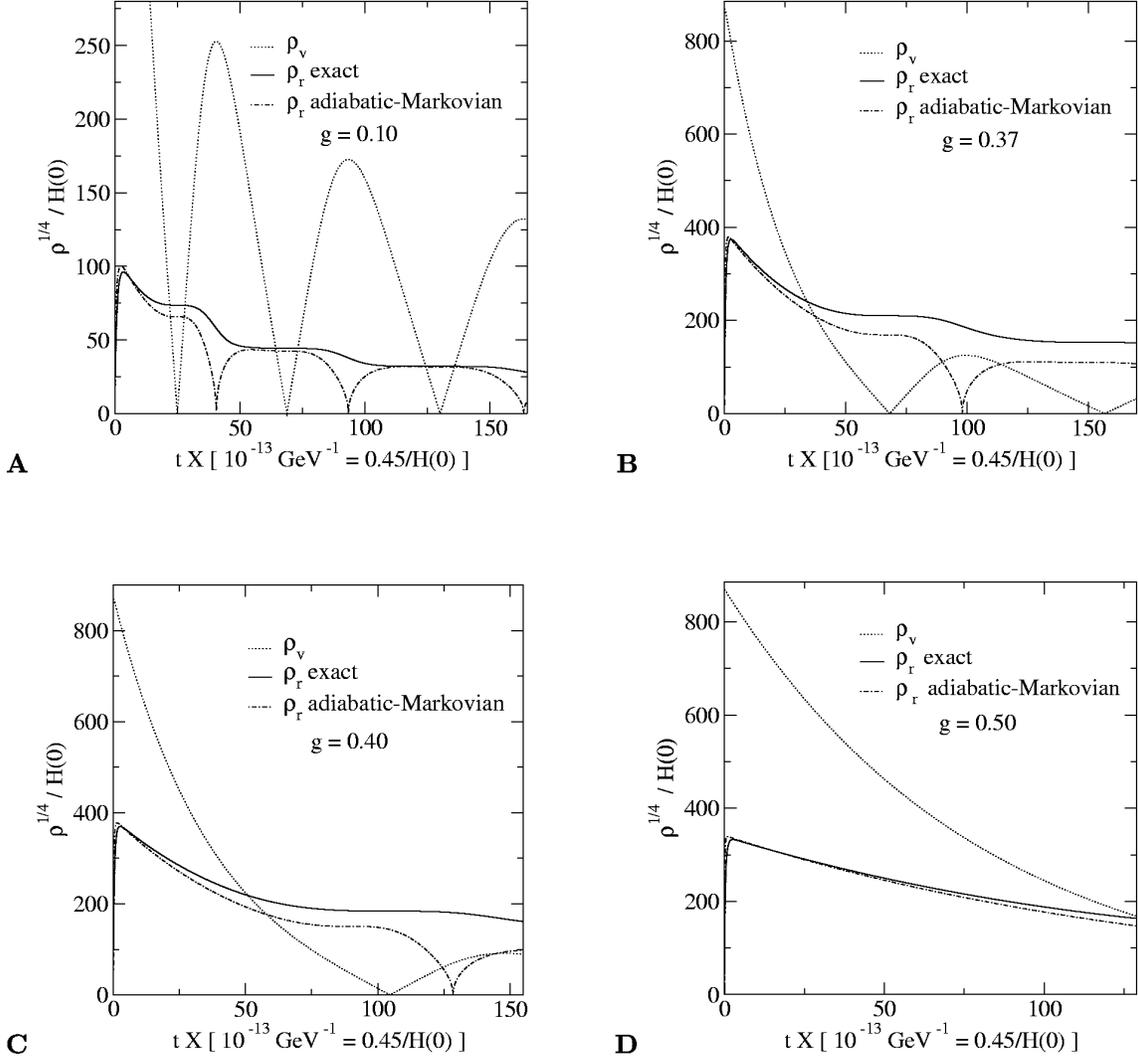}}}
\caption{Evolution of radiation density $\rho_r$, plotted
as the ratio $R \equiv \rho_r^{1/4}/H$ from various
approximations and various interaction couplings $g$,
with $h=g$, $\lambda=10^{-13}$,
$\xi=0$,
$\varphi(0) = m_{\rm Pl}$, and ${\dot \varphi}(0)=0$.}
\label{figrho}
\end{figure}

\begin{figure}[ht]
\vspace{1cm}
\epsfysize=6.5cm
{\centerline{\epsfbox{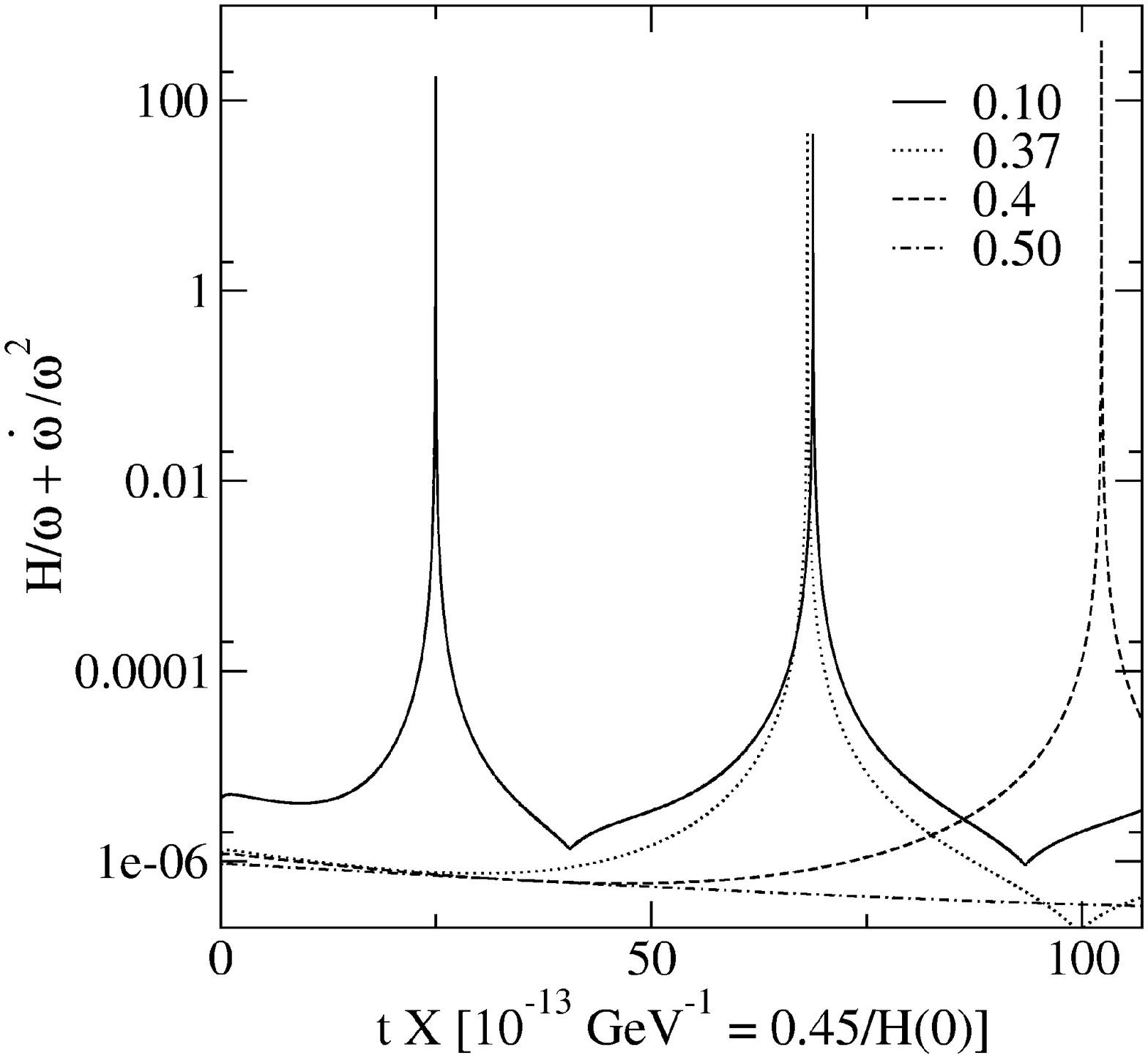}}}
\caption{Checking the adiabatic approximation 
Eqs. (\protect\ref{wkb cond}),
for various interaction couplings $g=0.1,0.37,0.4$ and $0.5$,
with $h=g$, $\lambda=10^{-13}$,
$\xi=0$,
$\varphi(0) = m_{\rm Pl}$, and ${\dot \varphi}(0)=0$.}
\label{figadtest}
\end{figure}

\subsection{$\varphi$ effective equation of motion}

The solutions for $\varphi(t)$ from the effective evolution equation are
plotted in {}Fig. \ref{figd01} for $g=0.1$ and {}Fig. \ref{figvp2}
frames a-c for $g=0.37,0.4$ and $0.5$, respectively. {}For each case the
solution is plotted from the exact one-loop evolution equation Eq.
(\ref{eom3a}) (solid), the same nonlocal evolution equation except with
the kernel being replaced with its nonexpanding space-time counterpart
(dashed), the adiabatic-Markovian evolution equation Eq.
(\ref{amapprox}) (dot-dashed) and the evolution equation where no
account for dissipative effects is treated (dotted). The latter dotted
curves are the ones assumed in cold inflation studies, where the effects
of dissipation are simply ignored. As Figs. \ref{figd01} and
\ref{figvp2} indicate, this assumption can be critically wrong. In
particular, for the large coupling cases in Fig. \ref{figvp2}, one
sees that the effect of dissipation drastically affects the behavior of
$\varphi(t)$ from the underdamped evolution found in the dotted curves
to overdamped evolution once dissipative effects are properly accounted
for. Comparing the three curves in each frame which treat dissipative
effects at different levels of approximation, we see that they are all
in excellent agreement. In particular, the adiabatic-Markovian
approximation, which is based on the simplified evolution equation Eq.
(\ref{amapprox}), is in excellent agreement with the exact evolution
equation Eq. (\ref{eom3a}), where the nonlocal kernel is fully treated
numerically. Based on our examination of the kernels in {}Figs.
\ref{figk} and \ref{figki}, and the fact that the integrated kernels
rapidly converge in a time $t \ll 1/H$, these results for $\varphi(t)$
come as no surprise.  

For the $g=0.37$ and $0.4$ cases,
the longtime behavior appears to show oscillations as opposed
to a complete overdamped relaxation.  This is an artifact of
our approximation of treating the $\varphi$-dependent $\chi$ field mass as
fixed to the value of the field amplitude at the initial time $t_0$.
This is done since then we can compute the kernel once and for all,
before evolving $\varphi$ in Eq. (\ref{eom3a}). Allowing the $\chi$ mass
to vary would require the kernel to be recomputed at every step of the
evolution and that would be far too time consuming for this calculation
be be tractable. However by doing this simplification, it leads to the
nonlocal damping term depending on the field amplitude as $\varphi^2$ whereas itshould be $\varphi$. This means as $\varphi \rightarrow 0$, our
approximation causes the nonlocal term to go to zero faster than it
actually should and in particular faster than the curvature of the
potential, thus leading to the oscillations. Thus for $g=0.37$ and
$0.4$, the oscillations are simply an artifact of approximations used in
numerically computing the $\varphi$ effective evolution equation.

Turning to Fig. \ref{figd01} for $g=0.1$, we find that the effect of 
dissipation is not significant
enough to alter the evolution of $\varphi(t)$ by very much.
This is a weak dissipative regime, where in the adiabatic-Markovian
approximation $\Upsilon < 3H$.  However the effect of dissipation is
not entirely negligible.  The two inset boxes in this figure
close-up on $\varphi(t)$ at the two extrema.
They show that the amplitude of $\varphi(t)$ is slightly less
in the cases where dissipation is treated in comparison with the
no dissipation (dotted curve) case.  This indicates
that energy is being depleted from the $\varphi$-system
into radiation.  

\subsection{Radiation production}

In particular, Fig. \ref{figrho} shows the ratio
$\rho_r^{1/4}/H$ for the cases $g=0.1, 0.37, 0.4$ and
$0.5$ in a-d respectively.  In each frame there is a plot
of $\rho_r$ from the exact evolution equations,
Eqs. (\ref{eom3a}) and (\ref{drad}) (solid)
and based on the adiabatic-Markovian evolution 
equations, Eqs. (\ref{amapprox}) and (\ref{rhor})
(dashed).  In all cases, the results show the exact and
adiabatic-Markovian approximation are in good agreement.
In particular, not only in the strong dissipative
regime $\Upsilon > 3H$ but also in the weak dissipative regime
$\Upsilon < 3H$, the simple formula Eq. (\ref{rhor}) for
determining radiation production is valid.

\subsection{Equilibration and thermalization: the asymptotic long
time behavior}

So far we have not discussed the long time behavior of our results, in
particular the equilibration and thermalization of the radiation
produced by the dissipation mechanism discussed in this paper. Before
doing that it is useful to show whether and how our results,
particularly the numerical ones given above for both the dissipative
kernel and for the evolution of the $\varphi$ background field, can
compare to recent numerical results obtained from the full evolution of
fields far-from-equilibrium  \cite{berges} and therefore not restricted to
quasi-equilibrium conditions only. Though here we have restricted our
study of the dynamics for field configurations close to equilibrium and
that evolve adiabatically, as we will see, our study still is able to
capture many characteristics observed in the recent studies of the
dynamics of scalar fields.

Recently the authors in \cite{berges} have shown extensive
numerical solutions for the kinetic (two-point correlation) equations
for a scalar field in 1+1 D. Since these results seem also qualitatively
to apply to 3+1 D, it is useful to see whether there is any
similarity with the general characteristics observed for the dynamics,
for both kernel and background field, obtained here compared to those
obtained in \cite{berges}. In particular, the authors of
\cite{berges} have shown that the dynamics of correlations (that also
applies to the time dependent number density evolution) can generically
be divided into three basic regimes: a damping regime characterized by
an exponential suppression of the correlations in a time scale $t_{\rm
damp} \sim 1/\Gamma$, followed by a drifting like evolution behavior
characterized by smooth and slow changing of the modes, which typically
lasts much longer than the initial damping evolution, after which a last
regime sets in, the thermalization itself, in which thermal equilibrium
is achieved, within a time scale $t_{\rm thermal} \gg t_{\rm damp}$. In
the kinetic (or Boltzmann-like) approach in which the time evolution of
correlations are solved, thermalization is seen as a direct consequence
of self-consistently including scattering processes in the kinetic
equations \cite{berges,jakovac,hu2}. In our case, this would be
equivalent to self-consistently take into account in our evolution
equation and in the derivation of the nonequilibrium propagators, the
backreaction of the produced radiation. This is fundamental in order to
describe the thermalization process, since in this way proper
equipartition of energy among the modes is taken into account and that
will then lead to thermal equilibration in the long time evolution of
the system field. Therefore, our results will not account for the very
long time thermalization regime. 
On the other hand, we can check from
our numerical results, in particular for the temporal behavior
for both the nonlocal dissipative kernel and also for its time
integrated form, Figs. {\ref{figk} and \ref{figki}, respectively, that
they show similar behavior to the correlations obtained by the authors
in Ref. \cite{berges}.
In particular Figs. {\ref{figk} and \ref{figki} show for the
kernel a quick damping of oscillations within the time
scale $1/\Gamma_\chi$, which is then followed by
a drift like behavior of almost unperturbed, small amplitude oscillatory
like evolution over a much larger time scale than
$1/\Gamma_\chi$.
Up to the time scales we have studied the dynamics, we have seen no
appreciable change
in this behavior.  Thus
the way our kernel evolves with these two regimes is
analogous to those observed in the
full kinetic equations approach of \cite{berges}.
This gives us an indication that, even though
backreaction due to radiation production is not being fully taken into
account, we are still capturing the relevant dynamics from the initial
time of evolution (of no radiation) up to some very long time scale. 
{}Furthermore, it can be checked from the results shown in Fig. 
\ref{figrho} that during the time scale of evolution that we have studied,
the produced radiation maintains a level that is only a fraction 
of the inflaton energy density, $\rho_r/\rho_\varphi \lesssim 10^{-5}$,
and, therefore, we expect its overall effect back on the evolution of the 
inflaton field to be only marginal. 

The full inclusion of scattering and then the description of the final,
asymptotic equilibration and thermalization regime, could in principle
be done within the kinetic, Boltzmann-like approach of
\cite{berges,jakovac,hu2}, or within other equivalent approaches able to
describe the thermalization and equilibration process such as
\cite{koide,boyareheat}. The extension of these approaches to
our expanding space-time multifield setting is an
interesting (and likely most more complex) avenue for future work, but it is
beyond the scope of this paper.

\subsection{Adiabatic and WKB approximations}

{}Finally we come to the analysis of some of our basic approximations
used to derive the $\varphi$-EOM in its different forms. In Fig.
\ref{figadtest} the validity of the adiabatic approximation is examined.
{}From Eq. (\ref{wkb cond}), recall this approximation is valid when
$H/{\omega} + {\dot \omega}/{\omega^2} \ll 1$ (y-axis in the figure).
{}For $g=0.5$, Fig. \ref{figvp2} shows that $\varphi$ remains
overdamped throughout evolution and correspondingly Fig. \ref{figadtest}
shows the adiabatic approximation remains excellent. However for $g=0.1,
0.37$ and $0.4$ there are peaks crossing above one in Fig. \ref{figadtest},
thus meaning the adiabatic approximation breaks down in those regions.
In comparing to Figs. \ref{figd01} and \ref{figvp2}, all these peaks
correspond to when the $\varphi(t)$ evolution seizes to be overdamped
and it goes through a maxima or minima. In these underdamped regimes, in
any event the dissipative term has no significant influence on the
evolution of the system and so the breakdown of this approximation is of
little consequence. Moreover, in the context of inflation, for $g=0.37$
and $0.4$ these breakdown regimes first occur at very late stages near
the end of inflation, and so are not in a regime of interest for large
scale structure formation. Also as commented earlier, in order to
make our calculation tractable on the computer, we treated
the $\varphi$-dependent $\chi$ field mass as
fixed to the value of the field amplitude at the initial time $t_0$.
This was done since then we can compute the kernel once and for all,
before evolving $\varphi$ in Eq. (\ref{eom3a}), and thus
cutting computation time by well over an order of magnitude
and so bringing the computation time in the range of days as opposed
to weeks.
However by doing this simplification, it leads to the
nonlocal damping term depending on the field amplitude as $\varphi^2$ whereas it
should be $\varphi$. This means as $\varphi \rightarrow 0$, our
approximation causes the nonlocal term to go to zero faster than it
actually should and in particular faster than the curvature of the
potential, thus leading to the oscillations. Thus for $g=0.37$ and
$0.4$, the oscillations are simply an artifact of approximations used in
numerically computing the $\varphi$ effective evolution equation. {}For
$g=0.1$ the breakdown regimes of the adiabatic-Markovian approximation
are real, however at early times, $t \lesssim 50$ in the units
shown in {}Fig. \ref{figadtest}, this approximation is
excellent, so, for instance, results for radiation production at this
time are reliable.

\section{Dissipative Mechanism in Supersymmetry models}
\label{susy}

In the regimes where 
our dissipative mechanism is large enough
to  affect inflation,  the interaction couplings also are
significantly large to yield
radiative corrections that harm the flatness of the
inflaton effective potential.  As such supersymmetry is
needed, since it can cancel temporally local radiative
effects from Bose and Fermi sectors, thus 
almost completely preserve the tree
level potential.  
On the other hand, temporally nonlocal
radiative effects, such as those that lead to dissipation,
have very different space-time structure between Bose and Fermi sectors,
and so are not canceled by SUSY.

As discussed in Sec. \ref{Veff-section}, the $\psi_{\chi}$-fermions were
included to mimic the effect of SUSY  by cancelling the quantum
corrections from the $\chi$-bosons.  However the basic dissipative
mechanism we have been studying in this paper,
light boson (inflaton) $\rightarrow$ heavy boson $\rightarrow$
light fermions, can be realized in very simple SUSY models.
{}For example, in Ref. \cite{BR3} we proposed the
following model of two superfields $\Phi$ and $X$,

\begin{equation}
W= \frac{1}{3}\sqrt{\lambda} \Phi^3 + g \Phi X^2 + 4m X^2,
\label{susymodel}
\end{equation}
where $\Phi = \phi + \psi \theta + \theta^2 F$ and
$X = \chi+ \theta \psi_{\chi} + \theta^2 F_{\chi}$
are chiral superfields.  The field
$\phi$ will be identified as the inflaton in
this model with $\phi = \varphi +\sigma$
and $\langle \phi \rangle = \varphi$.
This is the simplest SUSY model in which the inflaton
has a monomial potential, in this case

\begin{equation}
V_0(\varphi) = \frac{\lambda}{4} \varphi^4, 
\end{equation}
and which includes the standard
reheating interaction term to an additional boson $g^2 \phi^2 \chi^2$.
When $\varphi \ne 0$ there is a nonzero vacuum energy and so
SUSY is broken.  This manifests in the splitting of 
masses between the $\chi$ and $\psi_{\chi}$ SUSY partners with in 
particular

\begin{eqnarray}
m_{\psi_{\chi}}^2 & = & \left[
2 g^2 \varphi^2 +
16\sqrt{2}mg \varphi + 64m^2 \right]
\nonumber \\
m_{\chi_1}^2 & = & \left[
\frac{1}{8} (g^2 + \frac{1}{2}\sqrt{\lambda} g) \varphi^2 +
\sqrt{2}mg \varphi + 4m^2 \right] = m^2_{\psi_{\chi}}
+ \sqrt{\lambda}g \varphi^2
\nonumber \\
m_{\chi_2}^2 & = &  \left[
\frac{1}{8} (g^2 - \frac{1}{2}\sqrt{\lambda} g)\varphi^2 +
\sqrt{2}mg \varphi + 4m^2 \right] = m^2_{\psi_{\chi}}
-  \sqrt{\lambda}g \varphi^2 .
\label{cpmass}
\end{eqnarray}
One can check that
the one loop zero temperature effective potential 
correction in this case is
not significant to alter the flatness of the tree level inflaton potential, 

\begin{equation}
V_1(\varphi) \approx \frac{9}{128 \pi^2}
\lambda g^2 \varphi^4 \left[\ln\frac{m^2_{\psi_{\chi}}}{m^2} -2
\right ] \ll V_0(\varphi) = \frac{\lambda}{4} \varphi^4.
\end{equation}

\noindent
The authors of Ref. \cite{moss2} have recently studied independently 
the corrections in the
model (\ref{susymodel}) including also the effect of finite temperature
and reached analogous conclusion for the quantum corrections,
that the $T=0$ and now also the thermal corrections
can be kept under control. 
On the other hand this model has the interaction structure
of the form Eq. (\ref{lint}), and so leads to the dissipative 
mechanism studied in this paper.  In particular, 
one of the Yukawa couplings of this model is
$4g \chi_i \psi_{\chi} \psi$.  Noting the mass splittings
in Eqs. (\ref{cpmass}), it means that the heavier $\chi$ boson,
$\chi_1$ can decay into a $\psi_{\chi}$ fermion and
an effectively massless inflatino $\psi$.
There will be a phase space suppression in this process due
to the closeness in masses of $\psi_{\chi}$ and $\chi_1$
so that the decay width now is

\begin{equation}
\Gamma_{\chi_1 \rightarrow \psi_{\chi},\psi} = 
\frac{g \lambda}{4\sqrt{2}\pi} \varphi ,
\end{equation}
and this leads to the dissipative coefficient being

\begin{equation}
\Upsilon = \frac{\sqrt{2} g^4 \lambda}{256 \pi^2 m_{\chi}} \varphi.
\end{equation}
So in this case in general 
$\Gamma_{\chi_1 \rightarrow \psi_{\chi},\psi}, \Upsilon < H$.
The radiation level,  ${\cal R}$, during inflation is found 
from Eq. (\ref{rhor}) to be

\begin{equation}
{\cal R} \equiv \frac{\rho_r^{1/4}}{H} \approx 0.03 \frac{g^{3/4}}{\lambda^{1/8}}
\left(\frac{m_{\rm Pl}}{\varphi}\right)^{7/4}.
\end{equation}
Thus for $\lambda = 10^{-13}$ and $\varphi = m_{\rm Pl}$,
${\cal R}>1$ arises for $g > 0.73$.  However for $\varphi$
much larger than $m_{\rm Pl}$, ${\cal R}>1$ requires $g > 1$.
Thus this model is not very robust in producing radiation during
inflation, but nevertheless the effect also is not negligible.

The radiation production during inflation 
in the model Eq. (\ref{susymodel})
can be greatly enhanced by adding some
light fermions into which the $\chi_i$-bosons can decay.
In any event, in a realistic particle physics model
the inflaton sector, such as the model Eq. (\ref{susymodel}),
would interact with other fields.
This could be done for example with an another superfield
$Y$ added to the superpotential Eq. (\ref{susymodel}) as
$h X Y Y/2$, which leads to the Yukawa interaction term
$h \chi \psi_Y \psi_Y$.  Provided
$2m_Y \ll m_{\chi}$, the $\chi$ decay width is unsuppressed
and in particular would be just Eq. (\ref{Gammachi}) with all
other subsequent expressions there also applicable here.
For this model we find in the strong dissipative regime
$\Upsilon > 3H$ that

\begin{equation}
{\cal R} \equiv \frac{\rho_r^{1/4}}{H} \approx 2.66 
\frac{1}{g^{3/4}h^{1/2} \lambda^{1/8}} 
\left(\frac{m_{\rm Pl}}{\varphi}\right)^{5/4}.
\end{equation}
{}For $\lambda = 10^{-13}$ and $\varphi = 5m_{\rm Pl}$, which in
cold inflation analysis of this model would be approximately where
the 60th e-fold of inflation occurs, we get ${\cal R} > 1$
for $g^{3/2} h > 0.09$.  In the weak dissipative
regime $\Upsilon < 3H$ we find 

\begin{equation}
{\cal R} \equiv \frac{\rho_r^{1/4}}{H} \approx 0.036
\frac{g^{3/4}h^{1/2}}{\lambda^{3/8}} 
\left(\frac{m_{\rm Pl}}{\varphi}\right)^{7/4},
\end{equation}
for which ${\cal R}>1$ and the weak dissipation condition hold in
the regime $0.00004 < g^{3/2} h < 0.09$.
Thus warm inflation is very robust in this model.

\section{Influence of dissipation on density perturbations}
\label{iden}

Provided that the radiation component present during inflation is bigger
than the inflaton mass, $\rho_r^{1/4} > m_{\phi}$, one should generally
expect that this radiation component will influence the fluctuations of
the inflaton. Since the typical mass of the inflaton is $\sim H$, this
amounts to the criteria already mentioned in the Introduction
$\rho_r^{1/4} > H$. Moreover, if one assumes thermalization, so that
$\rho_r^{1/4} \approx T$, the inflaton fluctuations are then thermal. In
this case the effect that the radiation component has on density
fluctuations can be explicitly computed. Although it is beyond the scope
of this paper to address the issue of thermalization, as a reasonable
guideline thermalization is expected provided the decay width
$\Gamma_{\chi} >H$. In this section, some examples of density
perturbations during the warm inflation regime will be presented and the
differences will be compared to the comparable results that would be
obtained under the assumption of cold inflation.

{}For either ground state or thermal fluctuations of the inflaton, the
density perturbations are obtained by the same expression
\cite{Guth:ec},

\begin{equation}
\delta_H = \frac{2}{5}\frac{H}{\dot \varphi} \delta \varphi\;.
\end{equation}

\noindent
In the warm inflation regime the fluctuations of the 
inflaton go in the strong dissipative regime as \cite{abadiab}

\begin{equation}
\delta \varphi^2 = \left(\frac{\pi}{4}\right)^{1/2}
\sqrt{H\Upsilon}T\;,\;\;\;\; 
{\rm for \; warm \; inflation} \hspace{0.1cm}(\Upsilon > 3H),
\hspace{0.2cm} T > m_{\phi},
\label{dpsd}
\end{equation}
and in the the weak dissipative regime as \cite{Berera:1995wh}

\begin{equation}
\delta \varphi^2 = \left(\frac{3\pi}{4}\right)^{1/2}  HT \;,\;\;\;\; 
{\rm for \; warm \; inflation} \hspace{0.1cm}(\Upsilon < 3H),
\hspace{0.2cm} T > m_{\phi}.
\label{dpwd}
\end{equation}
In contrast, for cold inflation, where inflaton
fluctuations are exclusively quantum \cite{Guth:ec},

\begin{equation}
\delta \varphi^2 =\frac{H^2}{(2\pi)^2}, \;\;\;\;\; 
{\rm for \; cold \; inflation}, \hspace{0.2cm} T < m_{\phi} .
\label{dpci}
\end{equation}
The associated spectral indices, $n_s$, for these three cases are
\cite{Hall:2003zp}

\begin{equation}
n_s-1 \equiv \frac{d\ln\delta_H^2}{d\ln k} =
\frac{1}{r} \left( -\frac{9}{4} \epsilon + \frac{3}{2} \eta
-\frac{9}{4} \beta \right),
\;\;\;\;\; 
{\rm for \; warm \; inflation} \hspace{0.1cm}(\Upsilon > 3H),
\hspace{0.2cm} T > m_{\phi},
\label{nswisd}
\end{equation}

\begin{equation}
n_s-1 =
\left( -\frac{17}{4} \epsilon + \frac{3}{2} \eta
-\frac{1}{4} \beta \right),
\;\;\;\;\; 
{\rm for \; warm \; inflation} \hspace{0.1cm}(\Upsilon < 3H),
\hspace{0.2cm} T > m_{\phi},
\label{nswiwd}
\end{equation}
and \cite{spectrum}

\begin{equation}
n_s-1 =
\left( -6 \epsilon + 2\eta \right),
\;\;\;\;\; 
{\rm for \; cold \; inflation},
\hspace{0.2cm} T < m_{\phi},
\label{nsci}
\end{equation}

\noindent
where $k$ is the wavenumber of the inflaton mode
and the slow-roll parameters are defined as 
$\epsilon \equiv m_p^2V'^2/(16\pi V^2)$,
$\eta \equiv m_p^2V''/(8\pi V)$,
and 
$\beta \equiv m_p^2 \Upsilon' V'/(8\pi \Upsilon V)$.

The spectral index will now be compared between the warm and cold inflation
cases for the $\lambda \phi^4/4$ potential. 
{}For cold inflation the result are well known to be 
$n_s - 1 = -3/N_3$ \cite{spectrum},
where $N_e$ denotes the number of e-folds of inflation. So at $N_e=60$,
for example, $n_s - 1 = -1/20$,
with the model parameters
$\lambda = 8 \times 10^{-14}$
and $\varphi_{60} = 4.37 m_{\rm Pl} $. These parameters
correspond to the amplitude of the density perturbations of $\delta_H
\approx 10^{-5}$. 

Turning to the warm inflation
case,  we now consider this potential coupled to
additional fields in the manner of the Lagrangian Eq. (\ref{Nfields})
and account for the effects of radiation on density perturbations
given by Eqs. (\ref{dpsd}) and (\ref{dpwd}).  In this case for the same model
parameters and this same value $\varphi_{60}$ of the field amplitude,
these thermal effects increase the density perturbation normalization
in the strong dissipative regime to $1 \times 10^{-3}$.
%cases a and b respectively in the weak dissipative regime to $\delta_H >
%2 \times 10^{-4}$ and $5 \times 10^{-5}$ and in the strong dissipative
%regime $\delta_H > 6 \times 10^{-3}$ and $1 \times 10^{-3}$. Thus the
Thus the effect of dissipation and radiation production during inflation lead to
a noticeable change in the behavior of the inflaton and its fluctuations.
We now readjust the model parameters to properly normalize
the density perturbations to the same value 
as before so that at $N_e \approx 60$,
$\delta_H \approx  10^{-5}$.
Nevertheless the spectral index will still differ.
In particular normalizing the density perturbations as before
requires the parameters in the strong dissipative regime to now be
$\lambda \sim 10^{-17}$ and the spectral index at 60-efolds
becomes $n_s - 1 = -1.5/N_e$.  So for $N_e=60$ this implies
in the strong dissipative warm inflation regime the
$\lambda \phi^4/4$ potential leads to $n_s-1 = 0.025$ which is
half the size of the correspond cold inflation result.
Since the recent CMB satellite experiments, WMAP and upcoming Planck, 
should be able to
discriminate spectral indices at the one percent level, the
difference between warm and cold inflation might be detectable.
This Section was simply illustrating some points regarding
density perturbation differences in warm versus cold
inflation.  A detailed analysis of
this issue will be presented in \cite{bb}.

\section{Conclusion}
\label{concl}

In this paper we have developed a formalism for treating dissipation in
quantum field theory models with slowly evolving backgrounds in an
expanding spacetime. The key steps for doing this were first computing
the real-time matrix of dressed expanding space-time two-point Green's
functions for the respective quantum fields in our system. The solution
of these Green's functions was obtained in a WKB approximation, which is
valid for slow moving evolution of the background fields and for fields
with masses much bigger than the Hubble scale. Having derived these
Green's functions, we then used a standard response theory approximation
approach for the derivation of the field averages appearing in the
effective evolution equation for the background component of a scalar
field $\varphi \equiv \langle \Phi \rangle$. The integration of the
quantum field fluctuations employed a nonperturbative resummation. The
resulting effective evolution equation for the background field
$\varphi$ showed dissipative features.

As seen, our dissipative formalism differs from those used in treatments
of reheating after inflation. In those cases, one is studying a fast
moving, oscillating background component, typically in a linear
relaxation (small field amplitude) regime. In contrast our analysis is
applicable for slowly moving background fields that do not oscillate and
in the nonlinear regime for the system field (the inflaton). As shown in
Sec. \ref{compare}, the basic physics that underlies the dissipation in
the two cases are markedly different. We have applied the nonlinear
dissipative mechanism developed here to the inflationary regime.
However, this same dissipation mechanism could also apply to preheating
scenarios (if they are allowed by the model and given set of
parameters), where the linearized, perturbative approximation for the
inflaton breaks down and nonlinear, nonperturbative effects, such as the
one studied in this paper, become important.

In addition to deriving the basic equations for our dissipative
formalism, a detailed numerical analysis was done. In particular, the
key quantity that our formalism determines is temporally nonlocal terms
that must be included in the $\varphi$ background field effective
evolution equation. In Sec. \ref{numerical} these nonlocal kernels were
numerically calculated and compared at various levels of approximations.
{}For instance, the $\varphi$ effective EOM was computed with the exact
one-loop expression in Eq. (\ref{eom3a}). A key question was the regime
of validity of the simplified adiabatic-Markovian approximation Eq.
(\ref{amapprox}) to the exact equation. In {}Figs. \ref{figd01} and
\ref{figvp2} these comparisons were made. In the regime where the WKB
self consistency conditions Eq. (\ref{wkb cond}), together with the
condition $\Gamma_{\chi_j} > H$ are satisfied, we found in {}Figs.
\ref{figd01} and \ref{figvp2} that the evolution equation computed from
the exact one-loop expression and from the adiabatic-Markovian
approximation agree very well within the region of parameters we have
concentrated our study. We also checked in Sec. \ref{numerical} the
radiation production from the $\varphi$ system that emerges through
dissipation. Once again the exact numerical treatment and the
adiabatic-Markovian approximation agreed very well in the same regimes
as for the evolution equation. These results are of great practical use,
since calculating the exact 
numerical solution to the effective evolution equation
is very time consuming on the computer, whereas the evolution equation
in the adiabatic-Markovian approximation can be analyzed analytically.

The immediate application of our dissipative formalism is to
inflationary cosmology, in particular to determine warm inflation
regimes and their properties. In Secs. \ref{susy} and \ref{iden} we
examined some consequences. In general there are two sorts of
qualitative effects that dissipation can have on the inflationary phase.
First, the evolution of the background field can be altered due to the
nonlocal terms. The most dramatic example of that was shown in {}Fig.
\ref{figvp2} where accounting for dissipation, $\varphi$ evolution was
overdamped, whereas if one simply ignored these effects, the evolution
would have the underdamped oscillatory behavior typically assumed in
cold inflation studies. In particular, in the larger perturbative
coupling regimes studied in {}Fig. \ref{figvp2}, the inflaton field
would never have a reheating phase (in the sense of a fast oscillatory
like regime for $\varphi$). It would simply relax to the minimum of the
potential monotonically, dissipating radiation along the way, thus
ending the inflation phase and initiating a radiation dominated phase.
Less dramatic to this, but down to much lower interaction couplings, even
though the background inflaton field evolution is not noticeably altered
from dissipative effects, radiation is still being produced during
inflation from conversion of vacuum energy. Although there are detailed
questions about thermalization, that are beyond the scope of this paper
to address, as a reasonable criteria when $\rho_r^{1/4} > H$ during
inflation, one should expect that this radiation component will
influence the inflaton fluctuations significantly from its
zero-temperature zero-point level. To gain a better understanding of the
extent that this radiation can influence the inflaton fluctuations, and
hence the primordial seeds of density fluctuations, in Sec. \ref{iden}
we computed the density perturbations when the radiation component is
accounted for, and compared that to the naive expectation when the
fluctuations are assumed to be zero-point ground state fluctuations.

The key result of this study when applied to inflation has been that
dissipative effects are predicted to occur during inflation in typical
inflation models. These effects alter the single picture of inflationary
dynamics assumed up to now, which we call cold inflation, into another
possibility which we call warm inflation. To make accurate predictions
from inflation models, which is now required for current high-precision
CMB measurements, these dissipative effects must be treated. Moreover,
dissipation effects can lead to some attractive theoretical consequences
in inflation models. {}For example, for those parameter regions feasible
to inflation and where the nonlinear and nonperturbative effects we
studied here can become important, the emergence of effective strong
dissipative phenomena are able, for instance, to sustain and drive
inflation longer than when these dynamical effects are neglected
\cite{ROR_pascos}. Several other results also can follow as a consequence
of the dissipative regimes studied here. 
In particular various problems, namely $\eta$
\cite{Berera:2004vm}, graceful exit \cite{wi}, quantum-to-classical
transition \cite{wi,Bellini:ki}, large inflaton amplitude
\cite{Berera:2004vm}, and aspects of initial conditions \cite{bg,ROR},
can be remedied simply by properly accounting for the dissipative
effects already in the model, rather than relying on additional
modifications to the model, as is often done.

\begin{acknowledgments}

The authors thank J. Berges, I. Lawrie, M. Wise and M. Sher for
valuable discussions.
AB was funded by the United Kingdom Particle Physics and 
Astronomy Research Council (PPARC) and 
ROR was supported by Conselho Nacional de Desenvolvimento 
Cient\'{\i}fico e Tecnol\'ogico (CNPq-Brazil) and {}Funda\c{c}\~ao
de Amparo \`a Pesquisa no Estado do Rio de Janeiro (FAPERJ).

\end{acknowledgments}

\appendix

\section{The effective Potential and Renormalization}

Consider the local terms (\ref{dveff-b}) appearing in the 
$\varphi$-effective EOM and associated to the field derivative of the
effective potential for $\varphi$,

\begin{eqnarray}
\frac{\partial V_{\rm eff} (\varphi,R)}{\partial \varphi} &=&
m_\phi^2 \varphi(t) + 
\frac{\lambda}{6} \varphi(t)^3  +\xi R(t) \varphi(t) \nonumber \\
&+& 
\frac{\lambda}{2} \varphi(t) \frac{1}{a(t)^{3}} \int \frac{d^3 q}{(2 \pi)^3} 
\frac{1}{2 \left[{\bf q}^2/a(t)^2 + 
m_\phi^2 + \frac{\lambda}{2} \varphi(t)^2 + (\xi-1/6) R(t) \right]^{1/2}}
\nonumber \\
&+& \sum_{j=1}^{N_{\chi}} g_j^2 \varphi (t)
\frac{1}{a(t)^{3}} \int \frac{d^3 q}{(2 \pi)^3} 
\frac{1}{2 \left[{\bf q}^2/a(t)^2 + 
m_{\chi_j}^2 + g_j^2 \varphi(t)^2 + (\xi-1/6) R(t) \right]^{1/2}}\;,
\label{dVeff}
\end{eqnarray}

\noindent
where we used Eqs. (\ref{Gphitt}) and (\ref{Gchitt}). The momentum integrals
in Eq. (\ref{dVeff}) are divergent and require appropriate renormalization, which
we perform here just as in standard Minkowski space-time. In Eq. (\ref{dVeff})
we add mass and couplings renormalization counterterms to the classical potential
so as 

\begin{eqnarray}
\frac{\partial V_{\rm eff} (\varphi,R)}{\partial \varphi} \to
\frac{\partial V_{\rm eff} (\varphi,R)}{\partial \varphi} +
\delta m_\phi^2 \varphi(t) + 
\frac{\delta \lambda}{6} \varphi(t)^3  +\delta \xi R(t) \varphi(t)\;,
\end{eqnarray}
and we will consider from now on the masses and coupling constants as being
the renormalized ones. The counterterms $\delta m_\phi^2$, $\delta \xi$ and
$\delta \lambda$ are fixed by the choice of renormalization conditions 
\cite{ringwald,ringwald2})

\begin{eqnarray}
&&\frac{\partial^2 V_{\rm eff} (\varphi,R)}{\partial \varphi^2}\Bigr|_{\varphi=0,R=0}
=m_\phi^2  \;, \nonumber \\
&& \frac{\partial^3 V_{\rm eff} (\varphi,R)}{\partial R\partial \varphi^2}
\Bigr|_{\varphi=0,R=\mu_R^2}
= \xi  \;, \nonumber \\
&& \frac{\partial^4 V_{\rm eff} (\varphi,R)}{\partial \varphi^4}
\Bigr|_{\varphi=\mu_\varphi,R=0}
= \lambda \;,
\label{renorcond}
\end{eqnarray}

\noindent
where we have chosen renormalization points $\varphi=\mu_\varphi$ and
$R=\mu_R^2$ in the above conditions so that the results are infrared finite
in the limit of vanishing masses $m_\phi$ and $m_{\chi_j}$. Of course,
these renormalization points are completely arbitrary and related to 
different choices by the corresponding renormalization group equations.

Using an upper momentum cutoff $\Lambda$ and changing to the physical momentum
$k_p = k/a$ and cutoff $\Lambda_p = \Lambda/a$, the momentum integrals in
Eq. (\ref{dVeff}) are easily evaluated leading to (for $\Lambda \to \infty$)

\begin{eqnarray}
\frac{\partial V_{\rm eff} (\varphi,R)}{\partial \varphi} &=&
(m_\phi^2+\delta m_\phi^2) \varphi(t) + 
\frac{(\lambda+ \delta \lambda)}{6} \varphi(t)^3  +(\xi+\delta \xi) R(t) \varphi(t)
\nonumber \\
&+& \left[ \frac{\lambda}{2} + \sum_j g_j^2\right] \frac{\varphi}{8 \pi^2} \Lambda_p^2 
+ \frac{\lambda \varphi}{32 \pi^2} \left[ m_\phi^2 + \frac{\lambda}{2} \varphi^2 +
\left(\xi - \frac{1}{6} \right) R \right] 
\left\{1 + \ln \left[ \frac{m_\phi^2 +\frac{\lambda}{2} \varphi^2 + 
\left( \xi - \frac{1}{6}\right) R}{4 \Lambda_p^2} \right] \right\}
\nonumber \\
&+& 
\sum_j \frac{g_j^2 \varphi}{16 \pi^2} 
\left[ m_{\chi_j}^2 + g_j^2 \varphi^2 +
\left(\xi - \frac{1}{6} \right) R \right] 
\left\{1 + \ln \left[ \frac{m_{\chi_j}^2 +g_j^2 \varphi^2 + 
\left( \xi - \frac{1}{6}\right) R}{4 \Lambda_p^2} \right] \right\}
\;.
\label{dVeff2}
\end{eqnarray}
We can now use the renormalization conditions (\ref{renorcond}) in
(\ref{dVeff2}) leading, for massless bare fields 
$m_\phi=m_{\chi_j}=0$, to the renormalized expression

\begin{eqnarray}
\frac{\partial V_{\rm eff}^r (\varphi,R)}{\partial \varphi} &=& 
\frac{\lambda}{6} \varphi(t)^3  +\xi R(t) \varphi(t) 
-\frac{\lambda}{32 \pi^2} \left(\xi - \frac{1}{6}\right) R \varphi
\left\{1+ \ln \left[ \frac{ \left(\xi - \frac{1}{6}\right) \mu_R^2}{
\frac{\lambda}{2} \varphi^2 + \left(\xi - \frac{1}{6}\right) R} \right]
\right\}
\nonumber \\
&-& \sum_j \frac{g_j^2}{16 \pi^2} \left(\xi - \frac{1}{6}\right) R \varphi
\left\{ 1+ \ln \left[ \frac{ \left(\xi - \frac{1}{6}\right) \mu_R^2}{
g_j^2 \varphi^2 + \left(\xi - \frac{1}{6}\right) R} \right]
\right\} \nonumber \\
&-& \frac{\lambda^2}{64 \pi^2}  \varphi^3
\left\{\frac{11}{3}+ \ln \left[ \frac{\frac{\lambda}{2} \mu_\varphi^2}{
\frac{\lambda}{2} \varphi^2 + \left(\xi - \frac{1}{6}\right) R} \right]
\right\}
\nonumber \\
&-& \sum_j \frac{g_j^4}{16 \pi^2}  \varphi^3
\left\{\frac{11}{3}+ \ln \left[ \frac{g_j^2 \mu_\varphi^2}{
g_j^2 \varphi^2 + \left(\xi - \frac{1}{6}\right) R} \right]
\right\} \;.
\label{Vrenor}
\end{eqnarray}

We can also extend the result for the renormalized effective potential
when there is an additional coupling of $\Phi$ to fermions $\psi_\chi$,
in which case there is the additional contribution to (\ref{dVeff})

\begin{equation}
\frac{\partial V_{\rm eff} (\varphi,R)}{\partial \varphi} \to 
\frac{\partial V_{\rm eff} (\varphi,R)}{\partial \varphi} 
-\sum_k 4 h_k (m_\psi + h_k \varphi) \int \frac{d^3 k_p}{(2 \pi)^3}
\frac{1}{2 \sqrt{{\bf k}^2 + (m_\psi + h_k \varphi)^2}}\;,
\end{equation}

\noindent
which leads to the additional contribution to
(\ref{Vrenor}) (for bare massless $\psi_\chi$ fermions, $m_\psi=0$) 

\begin{equation}
\frac{\partial V_{\rm eff}^r (\varphi,R)}{\partial \varphi} \to 
\frac{\partial V_{\rm eff}^r (\varphi,R)}{\partial \varphi} +
\sum_k \frac{h_k^4 \varphi^3}{4 \pi^2} \left(
\frac{11}{3} + \ln \frac{\mu_\varphi^2}{\varphi^2} \right) \;.
\end{equation}


\begin{thebibliography}{99}

\bibitem{BGR} A. Berera, M. Gleiser and R. O. Ramos, Phys. Rev. {\bf D58}, 
123508 (1998).  
 
\bibitem{BGR2} A. Berera, M. Gleiser and R. O. Ramos, Phys. Rev. Lett. 
{\bf 83}, 264 (1999).  

\bibitem{BR}A. Berera and R. O. Ramos, Phys.
Rev. D{\bf 63}, 103509 (2001).

\bibitem{BR2}A. Berera and R. O. Ramos, Phys. Lett. B{\bf 567}, 294 (2003).

\bibitem{BR3}A. Berera and R. O. Ramos, In Press Physics
Letters B, arxiv: hep-ph/0308211, (2005).

\bibitem{oldi} A. H. Guth, Phys. Rev {\bf D23}, 347 (1981);
K. Sato, Phys. Lett. B{\bf 99}, 66 (1981).
                                                                                
\bibitem{ni} A. Albrecht and P. J. Steinhardt, Phys. Rev. Lett.
{\bf 48}, 1220 (1982); A. Linde, Phys. Lett. {\bf 108B}, 389 (1982).
                                                                                
\bibitem{ci} A. Linde, Phys. Lett. {\bf 129B}, 177 (1983).

\bibitem{wi} A. Berera,  Phys. Rev. Lett. {\bf 75},
3218 (1995);
Phys. Rev. {\bf D54}, 2519 (1996);
Phys.\ Rev.\  {\bf D55}, 3346 (1997).

\bibitem{HR}H. P. de Oliveira and  R. O. Ramos, 
Phys. Rev. {\bf D57}, 741 (1998); 
W. Lee and  L.-Z. Fang, Phys. Rev. {\bf D59}, 083503 (1999). 

\bibitem{old}L. F. Abbott, E. Fahri and M. Wise, Phys. Lett. {\bf 117B}, 29 (1982).

\bibitem{reheat} L. Kofman, A. Linde and A. A. Starobinsky,
Phys. Rev. D {\bf 56}, 3258 (1997); P. B. Greene and L. Kofman,
Phys. Lett. B{\bf 448}, 6 (1999); F. Finelli and R. Brandenberger,
Phys. Rev. D{\bf62}, 083502 (2000);


\bibitem{weldon}H. A. Weldon, Phys. Rev. {\bf D 28}, 2007 (1983).

\bibitem{weiss}U. Weiss, {\it Quantum Dissipative Systems} (World
Scientific, Singapure, 1999).
 
\bibitem{boya}D. Boyanovsky, H. J. de Vega, R. Holman, D.-S. Lee and A.
Singh, Phys. Rev. {\bf D51}, 4419 (1995).

\bibitem{GR} M. Gleiser and R. O. Ramos, Phys. Rev. {\bf D50}, 2441 
(1994). 

\bibitem{mallik} N. Banerjee and S. Mallik, Phys. Rev. {\bf D45}, 701 (1992).

\bibitem{CTP} K. Chou, Z. Su, B. Hao and L. Yu, Phys. Rep.
{\bf 118}, 1 (1985);
N. P. Landsman and Ch. G. van Weert, Phys. Rep.
{\bf 145}, 141 (1987).

\bibitem{ian}I. D. Lawrie, J. Phys. {\bf A25}, 6493 (1992); 
Phys. Rev. {\bf D60}, 063510 (1999). 

\bibitem{hu}  S. A. Ramsey, B. L. Hu and A. M. Stylianopoulos,
Phys. Rev. {\bf D 57}, 6003 (1998).

\bibitem{davis} N. D. Birrel and P. C. W. Davis, {\it Quantum Fields in Curved
Space} (Cambridge University Press, Cambridge, England, 1982);
S. A. Fulling, {\it Aspects of Quantum Field Theory in
Curved Space Time} (Cambridge University Press, Cambridge, England, 1989).
 
\bibitem{semeno}G. Semenoff and N. Weiss, Phys. Rev. D{\bf 31}, 699 (1985).

\bibitem{mallik2}H. Leutwyler and S. Mallik, Ann. Phys. (NY) {\bf 205}, 1
(1991).

\bibitem{habib} S. Habib, Phys. Rev. D{\bf 46}, 2408 (1992).

\bibitem{wkbad}
R. H. Brandenberger, S. E. Joras and J. Martin, Phys. Rev. D{\bf 66}, 083514
(2002);
J. Martin and R. H. Brandenberger, Phys. Rev. D{\bf 67}, 083512 (2003).


\bibitem{ringwald}
A. Ringwald, Ann. Phys. (NY) {\bf 177}, 129 (1987). 

\bibitem{ian3}  I. D. Lawrie, Phys. Rev. D{\bf 67}, 045006 (2003).


\bibitem{berges1}G. Aarts and J. Berges, Phys. Rev. {\bf D 64}, 
105010 (2001).

\bibitem{fetter}A. L. Fetter and J. D. Walecka, {\it Quantum 
Theory of Many Particle Systems} (McGraw-Hill, New York, 1971).

\bibitem{ringwald2}A. Ringwald, Phys. Rev. {\bf D36}, 2598 (1987).

%\cite{Vilenkin:sg}
\bibitem{Vilenkin:sg}
A.~Vilenkin,
%``Phase Transitions In De Sitter Space,''
Nucl.\ Phys.\ B {\bf 226}, 504 (1983).
%%CITATION = NUPHA,B226,504;%%

\bibitem{boya1} D. Boyanovsky, R. Holman and S. Prem Kumar,
Phys. Rev. {\bf D56}, 1958 (1997);
D. Boyanvsky and H. J. de Vega, arXiv: astro-ph/0406287.

\bibitem{berges} J. Berges and J. Cox,
Phys. Lett. {\bf B517}, 369 (2001);
J. Berges, Nucl. Phys. {\bf A699}, 847 (2002).

\bibitem{jakovac}A. Jakovac, Phys. Rev. {\bf D 65}, 085029 (2002).

\bibitem{hu2}E. Calzetta and B. L. Hu, Phys. Rev. {\bf D 61}, 025012 (2000).

\bibitem{koide} T. Koide and M. Maruyama, Nucl. Phys. {\bf A742}, 95 (2004);
T. Koide, Prog. Theor. Phys. {\bf 107}, 525 (2002).

\bibitem{boyareheat} D. Boyanovsky, M. D'attanasio, H. J. de Vega, R. Holman
and D.-S. Lee, Phys. Rev. {\bf D 52}, 6805 (1995);
D. Boyanovsky, I. D. Lawrie, D. S. Lee,
Phys. Rev. {\bf D 54}, 4013 (1996).

\bibitem{moss2} L. M. H. Hall and I. G. Moss, arXiv: hep-ph/0408323.

%\cite{Guth:ec}
\bibitem{Guth:ec}
A.~H.~Guth and S.~Y.~Pi,
%``Fluctuations In The New Inflationary Universe,''
Phys.\ Rev.\ Lett.\  {\bf 49}, 1110 (1982).
%%CITATION = PRLTA,49,1110;%%
 
\bibitem{abadiab} A. Berera, Nucl. Phys. {\bf B585}, 
666 (2000). 

%\cite{Berera:1995wh}
\bibitem{Berera:1995wh}
A.~Berera and L.~Z.~Fang,
%``Thermally induced density perturbations in the inflation era,''
Phys.\ Rev.\ Lett.\  {\bf 74},  1912 (1995).
%[arXiv:astro-ph/9501024].
%%CITATION = ASTRO-PH 9501024;%%

%\cite{Hall:2003zp}
\bibitem{Hall:2003zp}
L.~M.~H.~Hall, I.~G.~Moss and A.~Berera,
%``Scalar perturbation spectra from warm inflation,''
Phys.\ Rev.\ D {\bf 69}, 083525 (2004).
%%CITATION = ASTRO-PH 0305015;%%

\bibitem{spectrum} A. R. Liddle and D. H. Lyth, Phys. Lett. B {\bf 291},
391 (1992).

\bibitem{bb} M. Bastero-Gil and A. Berera, In Press
Physical Review D, hep-ph/0411144, (2005). 

\bibitem{ROR_pascos} R. O. Ramos, arXiv: hep-ph/0409353.

%\cite{Berera:2004vm}
\bibitem{Berera:2004vm}
A.~Berera,
%``Warm inflation solution to the eta-problem,''
arXiv:hep-ph/0401139.
%%CITATION = HEP-PH 0401139;%%

%\cite{Bellini:ki}
\bibitem{Bellini:ki}
M.~Bellini,
%``Warm Inflation And Classicality Conditions,''
Phys.\ Lett.\ B {\bf 428}, 31 (1998).
%%CITATION = PHLTA,B428,31;%%

\bibitem{bg} A. Berera and C. Gordon, Phys. Rev. D{\bf 63}, 063505
(2001).

\bibitem{ROR} R. O. Ramos, Phys. Rev. D{\bf 64}, 123510 (2001).


\end{thebibliography}
\end{document}